\documentclass[aps,prd,groupedaddress,showpacs,nofootinbib,amssymb,balancelastpage,preprintnumbers,floatfix]{revtex4}

\usepackage[dvipdfmx]{graphicx}
\usepackage{graphicx,bm,color}

\usepackage{amsmath}
\usepackage{amssymb}
\usepackage{amsfonts}
\usepackage{cases}
\usepackage{cancel}
\usepackage{hyperref}
\usepackage[normalem]{ulem}
\usepackage{subfigure}
\usepackage{here}
\usepackage{color}
\usepackage{comment}

\usepackage{tikz}
\usetikzlibrary{matrix}

\allowdisplaybreaks[3]

\begin{document}
	

\title{Impact of local CP-odd domain in hot QCD 
on axionic domain-wall interpretation \\ 
for NANOGrav 15-year Data}

\author{Linlin Huang}\thanks{{\tt huangll22@mails.jlu.edu.cn}}
\affiliation{Center for Theoretical Physics and College of Physics, Jilin University, Changchun, 130012,
	China}

\author{Yuanyuan Wang}\thanks{{\tt yuanyuanw23@jlu.edu.cn}}
\affiliation{Center for Theoretical Physics and College of Physics, Jilin University, Changchun, 130012,
China}

\author{He-Xu Zhang}\thanks{{\tt hxzhang18@163.com}}
\affiliation{Center for Theoretical Physics and College of Physics, Jilin University, Changchun, 130012,
	China}

\author{Hiroyuki Ishida}\thanks{{\tt ishidah@pu-toyama.ac.jp}}
	\affiliation{Center for Liberal Arts and Sciences, Toyama Prefectural University, Toyama 939-0398, Japan}

\author{Mamiya Kawaguchi}\thanks{{\tt mamiya@ucas.ac.cn}} 
      \affiliation{ 
School of Nuclear Science and Technology, University of Chinese Academy of Sciences, Beijing 100049, China
}

\author{Shinya Matsuzaki}\thanks{{\tt synya@jlu.edu.cn}}
\affiliation{Center for Theoretical Physics and College of Physics, Jilin University, Changchun, 130012, China}%

\author{Akio Tomiya}\thanks{{\tt akio@yukawa.kyoto-u.ac.jp}}
\affiliation{
Department of Information Technology, International Professional University of Technology in Osaka, 3-3-1 Umeda, Kita-Ku, Osaka, 530-0001, Japan
}

\begin{abstract}
We argue that the axionic domain-wall with a QCD bias may be incompatible with 
the NANOGrav 15-year data on a stochastic gravitational wave (GW) background, 
when the domain wall network collapses in the hot-QCD induced local CP-odd domain. 
This is due to the drastic suppression of the QCD bias set by 
the QCD topological susceptibility in the presence of the CP-odd domain 
with nonzero $\theta$ parameter of order one which the QCD sphaleron could generate. 
We quantify the effect on the GW signals 
by working on a low-energy effective model of Nambu-Jona-Lasinio type 
in the mean field approximation. 
We find that only at $\theta=\pi$, the QCD bias tends to get 
significantly large enough due to the criticality of the thermal CP restoration, 
which would, however, give too big signal strengths to be consistent with 
the NANOGrav 15-year data and would also 
be subject to the strength of the phase transition at the criticality. 
\end{abstract}

\maketitle


\section{Introduction} 

The observation of a stochastic GW background
has recently reported from the NANOGrav pulsar timing array
collaboration in 15 years of data~\cite{NANOGrav:2023gor,NANOGrav:2023hfp}. 
Possible origins of the detected nano-Hz peak frequency 
in a view of particle physics have been 
investigated also by the NANOGrav collaboration~\cite{NANOGrav:2023hvm}. 
Other recent pulsar timing array (PTA) data, such as those 
from the European PTA (EPTA)~\cite{EPTA:2023sfo,EPTA:2023akd,EPTA:2023fyk}, 
Parkes PTA (PPTA)~\cite{Reardon:2023gzh,Reardon:2023zen}, and Chinese PTA (CPTA)~\cite{Xu:2023wog} have 
also supported the presence of consistent 
nano-Hz stochastic GWs.   
Thus this nano-Hz GW evidence might provide us with 
a hint on the new aspect of the thermal history of the universe in terms of 
beyond the standard model of particle physics.

Among the new-physics candidate interpretations, 
the axionlike particle (ALP)-domain wall annihilation triggered by a QCD-induced bias 
has been considered as an attractive model%
~\cite{Li:2024psa,Ellis:2023oxs,Lozanov:2023rcd,Gelmini:2023kvo,Kitajima:2023cek,Geller:2023shn,Bai:2023cqj,Blasi:2023sej}, 
which can naturally be realized in the QCD-phase transition epoch 
at the temperature $T\sim {\cal O}$(100) MeV consistent 
with the produced peak frequency of nano Hz~\cite{NANOGrav:2023hvm}. 
The QCD-induced bias is supplied by the topological susceptibility $\chi_{\rm top}$, 
and its $T$-dependence and the value at $T=0$ have already been measured in 
a lattice simulation at the physical point for quark masses with 
the continuum limit is properly taken~\cite{Borsanyi:2016ksw}. 
This is how the ALP-domain wall prediction gets definite and unambiguous 
except the domain-wall network formulation and annihilation analysis, even though the system that the ALP 
acts in is nonperturbative QCD.

However, the thermal history of the QCD phase transition epoch may not be so simple: 
a local CP-odd domain may be created in hot QCD plasma 
due to the presence of the QCD sphaleron~\cite{Manton:1983nd,Klinkhamer:1984di}, so that the QCD vacuum characterized by 
the strong CP phase $\theta$ and its fluctuation (in the spatial-homogeneous direction) gets significantly sizable~\cite{Kharzeev:2007tn,Kharzeev:2007jp,Fukushima:2008xe} within the QCD time scale~\cite{McLerran:1990de,Moore:1997im,Moore:1999fs,Bodeker:1999gx}~\footnote{
The QCD sphaleron transition rate is not suppressed by the thermal effect in contrast to 
the QCD instanton's one~\cite{Moore:1997im,Moore:1999fs,Bodeker:1999gx}. 
The topological charge fluctuation (within the QCD time scale $={\cal O}(1 \mathchar`-10) $ fm), 
$\Delta q_{\rm top}(t) = \int^t dt' d \vec{x}^3 Q_{\rm top}(t', \vec{x})$, will be non-vanishing. 
This implies that the corresponding source $\theta(x)$ in the generating functional $Z_{\rm QCD}$ needs to be (at least) time-dependent and fluctuate: $\delta Z_{\rm QCD}/\delta \Delta \theta(t) \sim q_{\rm top}(t) \neq 0$. The time and/or thermal average of $\Delta \theta(t)$ 
thus acts as what we call the theta parameter 
in the QCD thermal plasma, which namely means $\theta \equiv \Delta \theta(t)$ in there. 
See also, e.g., the literature~\cite{Andrianov:2012hq,Andrianov:2012dj}. 
Besides, the time fluctuation $\partial_t \theta(t)$, to be referred to as the chiral chemical potential~\cite{Kharzeev:2007tn,Kharzeev:2007jp,Fukushima:2008xe,Andrianov:2012hq,Andrianov:2012dj}, will be significant as well when the non-conservation law of the $U(1)$ axial symmetry is addressed. 
See also Summary and Discussions. 
}. Though $\theta$ should be tiny enough ($< 10^{-10}$) at present, 
nonzero $\theta$ contribution to the QCD bias $\chi_{\rm top}$ 
may be non-negligible when the ALP domain wall starts to collapse in 
the QCD phase transition epoch. 
This is irrespective to whether or not the ALP acts as the QCD axion which relaxes the 
$\theta$ to zero at present.

In this paper, 
we argue that the ALP-domain wall 
with the QCD bias becomes incompatible with the NANOGrav 15-year data on a stochastic GW background, 
when the domain wall network collapses in the hot-QCD induced local-CP odd domain. 
This happens due to the drastic suppression of the QCD bias set by 
the QCD topological susceptibility in the presence of the CP-odd domain 
with nonzero $\theta$ parameter of order one which the QCD sphaleron could generate.

This paper is structured as follows. 
We first make a generic argument. It is based only on 
the anomalous Ward-Takahashi identities in QCD and 
the mixing structure between the scalar quark condensate ($\langle \bar{q}q \rangle$)  
and pseudoscalar quark condensate ($\langle \bar{q} i \gamma_5 q \rangle$) bilinear operators 
via the $U(1)$ axial transformation with nonzero $\theta$. 
$\chi_{\rm top}$ is shown to generically get small when $\theta$ takes the value of order one, 
because of the dramatic suppression of the scalar quark condensate at any temperature.

We next implement this suppression effect into the produced GW signals 
by working on a low-energy effective model of Nambu-Jona-Lasinio (NJL) type 
in the mean field approximation (MFA) and the random phase approximation (RPA). 
In accord with the generic argument, $\chi_{\rm top}$ 
as well as the scalar quark condensate get highly suppressed when $\theta = {\cal O}(1)$, 
instead, the pseudoscalar condensate as the CP/axial partner develops. 

We also find that only at $\theta=\pi$, the QCD bias tends to get 
significantly large enough due to the criticality of the thermal CP restoration of the second order, which is signaled when the pseudoscalar 
condensate reaches zero. 
This, however, gives too big signal strengths to be consistent with 
the NANOGrav 15-year data. 
Going beyond the MFA or other types of effective models could yield a different strength of the phase transition at the criticality. 
In summary of the present paper, we also briefly address the outlook 
along this possibility and the prospective impact on the QCD bias for the ALP-domain wall annihilation in light of nano Hz GW signals.

\section{General argument}\label{Sec-II}

We begin with presenting a general consequence on the suppression of $\chi_{\rm top}$ when $\theta = {\cal O}(1)$. 
First of all, we note that the anomalous $U(1)$ axial transformation ($q \to e^{-i \gamma_5 \frac{\theta}{4}} q \equiv q' $) can make the $\theta$ dependence in QCD 
present only in the quark mass term. 
In two-flavor QCD, for simplicity, 
the quark mass term then takes the form 
\begin{align}
   \sum_{q=u,d} \left[    m \cos\frac{\theta}{2}  (\bar{q}'q')  
    + 
    m \sin \frac{\theta}{2} (\bar{q}' i \gamma_5 q') 
    \right] 
\,,  \label{q-mass-term}
\end{align}
where the primed quark bilinear fields are related to the original ones via 
the orthogonal rotation, 
\begin{align}
    \begin{pmatrix}
     \bar{q} q  \\ 
     \bar{q} i \gamma_5 q  
    \end{pmatrix}
    =
    \begin{pmatrix}
     \cos \frac{\theta}{2} &  \sin\frac{\theta}{2} \vspace{2mm}\\ 
     - \sin\frac{\theta}{2} & \cos\frac{\theta}{2} 
    \end{pmatrix}
    \begin{pmatrix}
     \bar{q}' q'  \\ 
     \bar{q}' i \gamma_5 q'   
    \end{pmatrix}    
    \,. \label{mixing}
\end{align}
The scalar condensate $\langle \bar{q}q \rangle$ thus mixes with the pseudoscalar 
one $\langle \bar{q} i \gamma_5 q \rangle$ by nonzero $\theta$.

$\chi_{\rm top}$ is given as the functional derivative of the generating functional of QCD 
twice with respect to $\theta$ evaluated at $\theta\neq 0$. 
Taking into account the quark mass term in Eq.(\ref{q-mass-term}) we thus get  
\begin{align} 
 i \chi_{\rm top}(T, \theta) &= \int_T d^4 x \langle  Q_{\rm top}(x) Q_{\rm top}(0) \rangle_{\theta}  
\,, \notag \\  
Q_{\rm top}(x) & = \frac{g_s^2}{32 \pi^2} G_{\mu\nu} \widetilde{G}^{\mu\nu}
\,, 
\end{align}
where $Q_{\rm top}$ denotes the topological charge with the QCD gauge coupling $g_s$ 
and the (dual) field strength $G_{\mu\nu}$ ($\tilde{G}_{\mu\nu} \equiv 
\frac{\epsilon_{\mu\nu\rho\sigma}}{2} G^{\rho \sigma}$);  
$\int_T d^4 x \equiv \int^{1/T}_0 d \tau \int d^3 x$ with the imaginary time $\tau = i x_0$; the subscript ``$\theta$" attached on the vacuum (thermodynamic ground) states 
stands for the implicit $\theta$ parameter dependence.  
Following the procedure in the literature~\cite{Kawaguchi:2020qvg,Cui:2021bqf,Cui:2022vsr} 
$\chi_{\rm top}$ in two-flavor QCD 
is expressed in terms of the original-basis fields as 
\begin{align} 
\chi_{\rm top}(T, \theta) 
= - \frac{1}{4} \left[ 
	m \sum_{q=u,d} \langle \bar{q} q \rangle_{\theta,T}  
	+ i m^2 \chi_\eta(T, \theta) 
	\right] 
 \,, \label{chitop} 
\end{align}
where we have taken the isospin symmetric mass for up ($u$) and down ($d$) quarks, and 
\begin{align} 
& \chi_{\eta}(T, \theta) 
    =
    \int_T d^4 x \sum_{q=u,d}
\langle (\bar q(0) i \gamma _5  q(0))
	(\bar q(x) i \gamma _5 q(x)) \rangle_\theta    
 %
	\,, \label{chi-eta}
\end{align}
which is $\sim \frac{\partial}{\partial T} \sum_{q=u,d} \langle \bar{q}i \gamma_5 q \rangle_{\theta,T}$. 
Since the quark mass is perturbatively small enough 
and $\langle \bar{q}q \rangle$ develops a nonzero value at the normal QCD vacuum with $\theta=0$, 
the chiral perturbation conventionally works well for $\chi_{\rm top}$, 
so that the $\langle \bar{q}q \rangle$ term will dominate over the $\chi_\eta$ term, 
to saturate the measured $\chi_{\rm top}$ value $\simeq (75.6\,{\rm MeV})^4$, i.e., 
\begin{align}
    |\chi_{\rm top}(T,\theta=0)| \approx \frac{1}{4} m \sum_{q=u,d} |\langle \bar{q}q \rangle_{\theta=0, T}| \simeq (75.6\,{\rm MeV})^4 
\,.   \label{chitop-LS}
\end{align}
This is sort of the well-known formula called the Leutwyler-Smiluga formula~\cite{Leutwyler:1992yt}, and this  
feature has also been observed in chiral 
effective model approaches at any $T$~\cite{Kawaguchi:2020kdl,Kawaguchi:2020qvg,Cui:2021bqf,Cui:2022vsr}. 
When $\theta \neq 0$, however, 
the value of $\langle \bar{q}q \rangle$ 
will be shared with the pseudoscalar condensate $\langle \bar{q} i \gamma_5 q \rangle$ 
through Eq.(\ref{mixing}), so that 
$\langle \bar{q}q \rangle$ term in $\chi_{\rm top}$ of Eq.(\ref{chitop}) gets small, 
to be as small as or smaller than the $\chi_\eta$ term, as $\theta$ gets sizable. 
Thus, we expect the dramatic suppression of $\chi_{\rm top}$ at a sizable $\theta$~\footnote{
Even when the strange quark contributions are incorporated in $\chi_{\rm top}$, 
the form of Eq.(\ref{chitop}) is intact, as has been discussed in the literature~\cite{Kawaguchi:2020kdl,Kawaguchi:2020qvg,Cui:2021bqf,Cui:2022vsr} and also reviewed in Appendix~\ref{ACWTIs}, hence 
$\chi_{\rm top}$ can be evaluated only via the lightest quark condensate 
and $\chi_\eta$, in which the strange-quark loop contributions are implicitly incorporated. Thus the lightest quark condensate term ($\sum_{q=u,d}\langle \bar{q}q \rangle$) in 
$ |\chi_{\rm top}(T,\theta = {\cal O}(1))|$ still persists being suppressed due to the sizable CP violation, while the $\chi_\eta$ term keeps sizable with the large CP violation, where 
the CP-violating strange quark contribution almost decouples simply because of its heaviness 
(see also Summary and Discussions). Therefore,  
the inequality in Eq.(\ref{chi-top-suppress}) holds even in the case of 
three-flavor QCD. 
}: 
\begin{align}
    |\chi_{\rm top}(T,\theta = {\cal O}(1))| \approx \frac{1}{4} m^2 |\chi_\eta(T, \theta)|
    \ll  |\chi_{\rm top}(T,\theta=0)| 
\,. \label{chi-top-suppress}
\end{align}
This trend will be seen for any $T$ and $\chi_{\rm top}$ would simply get smaller and 
smaller as $T$ increases unless a sharp phase transition in $\chi_{\eta}$ shows up at higher $T$. 
Thus, the magnitude of the QCD-induced bias for the ALP domain wall annihilation, 
controlled by $\chi_{\rm top}$, is expected to become dramatically small 
in the local CP-odd domain created in hot QCD.

Below we will explicitize this claim based on an NJL with nonzero $\theta$.

\section{NJL evaluation of $\chi_{\rm top}$ and ALP domain wall energy density with nonzero $\theta$} 

Since lattice data on $\chi_{\rm top}(T, \theta)$ with $\theta \neq 0$ have not yet been available~\footnote{For more on the current status and future prospects, see also Summary and Discussions.}, 
we employ a low-energy chiral effective model, an NJL model,  
which matches with the underlying QCD via the consistent anomalous Ward-Takahashi identities associated 
with the chiral $SU(2)$ and $U(1)$ axial symmetry breaking. 
The investigation of QCD with nonzero $\theta$ has so far been 
carried out based on several chiral effective models
~\cite{Dashen:1970et,Witten:1980sp,Pisarski:1996ne,Creutz:2003xu,Mizher:2008hf,Boer:2008ct,Creutz:2009kx,Boomsma:2009eh,Sakai:2011gs,Chatterjee:2011yz,Sasaki:2011cj,Sasaki:2013ewa,Aoki:2014moa,Mameda:2014cxa,Verbaarschot:2014upa,Bai:2023cqj} and also the recently developed `t Hooft-anomaly matching method~\cite{Gaiotto:2017tne,Gaiotto:2017yup} extended from the original idea~\cite{tHooft:1979rat,Frishman:1980dq}, so as to clarify the nature of the thermal chiral and strong CP phase transitions when $\theta=\pi$. 
Nevertheless, the $T$ and $\theta$ dependence on $\chi_{\rm top}$ has never been addressed  
except Ref.~\cite{Sasaki:2013ewa}.  
In the reference $\chi_{\rm top}$ was computed based on the same NJL model 
as what we will work on below, however, 
consistency with the anomalous chiral Ward-Takahashi identities was not manifest, 
which is to be refined in the present study. 
We leave all the technical details in Appendices~\ref{ACWTIs} and \ref{NJL-PNJL} and shall only give 
the results relevant to the discussion of the nano Hz GW signals.

\begin{figure}[t] 
\centering
	\includegraphics[width=0.55\linewidth]{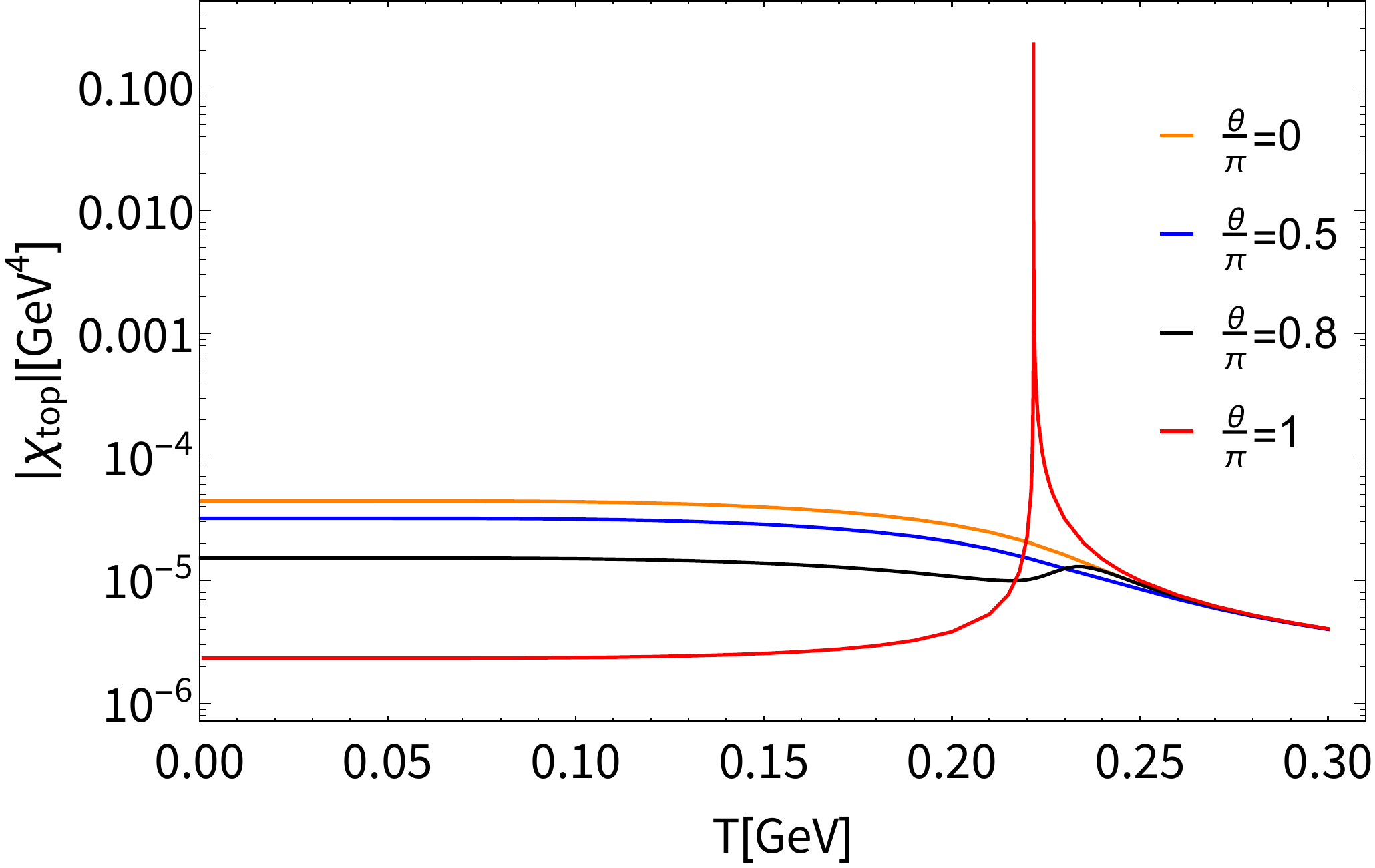}
    \caption{Plot of $\chi_{\rm top}$ (in magnitude) computed from the present two-flavor NJL model with nonzero $\theta$. 
 } 
 \label{chitop-fig}
\end{figure}

The dependence of $T$ and $\theta$ on $\chi_{\rm top}$ is plotted in 
Fig.~\ref{chitop-fig}. 
The present model yields 
$|\chi_{\rm top}^{1/4}(T=0, \theta=0)| \simeq 77.5$ MeV, which is in good agreement with  
the lattice estimate, $\chi^{1/4}_{\rm top}(T=0, \theta=0) \simeq 75.6$ MeV~\cite{Borsanyi:2016ksw}. 
As seen from the left panel of the figure, 
$\chi_{\rm top}$ prominently gets smaller when $\theta \gtrsim 0.5 \pi$, 
which is due to the suppression of $\langle \bar{q}q \rangle$ (as shown in Appendix~\ref{NJL-PNJL}), 
instead $\langle \bar{q} i \gamma_5 q \rangle$ gets greater than $\langle \bar{q}q \rangle$, 
as was expected from the generic argument in Eq.(\ref{chi-top-suppress}). 
We also note from Appendix~\ref{NJL-PNJL} that 
the chiral phase transition goes like crossover for any $\theta$, 
whereas the CP phase transition is of the second order type  
at $\theta = \pi$ (see Fig.~\ref{chitop-fig}), 
which is also manifested as a spike structure of $\chi_{\rm top}$. 
In particular, the CP symmetry is restored at the criticality ($T\simeq 221$ MeV) in accordance with the literature~\cite{Boer:2008ct,Boomsma:2009eh,Sakai:2011gs}.

\textcolor{black}{
We assume that an ALP has already been present before the QCD phase transition epoch 
as in the literature~\cite{Li:2024psa,Ellis:2023oxs,Lozanov:2023rcd,Gelmini:2023kvo,Kitajima:2023cek,Geller:2023shn,Bai:2023cqj,Blasi:2023sej} and developed the potential having the shift symmetry, 
$a \to a + 2 \pi f_a$ with $n_{\rm DW}$ being integer: 
\begin{align} 
 V_0(a) = \frac{m_a^2 f_a^2}{n_{\rm DW}^2} \left(1 - \cos \left(n_{\rm DW} \frac{a}{f_a} \right) \right)\,, 
\end{align}
where $m_a$ and $f_a$ are the ALP mass and decay constant, respectively. 
Then a domain wall profile exists as a soliton solution, 
which sweeps between the adjacent vacua, where 
$n_{\rm DW}$ in $V_0$ above corresponds to the domain wall number. 
As the universe cools down to the QCD scale, 
assuming existence of the ALP coupling 
to gluon fields, the ALP develops another 
potential via the $U(1)$ axial anomaly, 
\begin{align} 
 V_{\rm QCD}(a) = - \chi_{\rm top}(T, \theta)  \cos \left( n \frac{a}{f_a} + \theta \right)     
\,,   
\end{align}
which explicitly breaks the original shift symmetry,  
where $n$ is 
some factor related to and highly depending on the Peccei-Quinn charges of quarks. 
The original ALP vacuum-potential energy $\sim m_a^2 f_a^2$ is assumed to be $\gtrsim |\chi_{\rm top}|$. 
Thus the domain wall configuration, which is supported by the original shift symmetry, 
becomes unstable in the total ALP potential $V_{\rm total} = V_0 + V_{\rm QCD}$: 
\begin{align}
    V_{\rm total}(a) = \frac{m_a^2 f_a^2}{n_{\rm DW}^2} 
    \left( 1 - \cos\left( n_{\rm DW} \frac{a}{f_a} \right) \right) 
    + \chi_{\rm top} 
    \left(\cos\theta - \cos \left( n \frac{a}{f_a} + \theta \right) \right) 
    \,, 
\end{align}
with the normalization $V_{\rm total}(a=0)=0$. 
Meanwhile, the domain wall network (with $n_{\rm DW} \ge 2$) starts to collapse and release the latent heat into the universe.
Here, $V_{\rm QCD}(a)$ plays the role of what is called the bias, $\Delta V$. 
The released energy density can be evaluated by the vacuum energy at 
the original vacua $a = n_{\rm DW} \pi f_a$. For instance, when $n_{\rm DW}=2$,  
the original vacuum at $a/f_a= \pi$ gets the energy shift by $|V_{\rm total}(a/f_a=\pi)| 
= |V_0 - V_{\rm QCD}|_{a/f_a = \pi}$: 
\begin{align} 
|V_0 - V_{\rm QCD}|_{a/f_a = \pi}   
 = 
|\chi_{\rm top}(T, \theta) (\cos (n \pi + \theta) - \cos \theta) | \sim
|\chi_{\rm top}(T, \theta)| \equiv 
\Delta V(T, \theta) 
\,. 
\end{align} 
Thus one may maximally have the released latent heat $\rho_{\rm DW}(T)$ 
\begin{align} 
\rho_{\rm DW} \sim \Delta V(T, \theta) = |\chi_{\rm top}(T, \theta)|
\,.  
\end{align} 
Similar discussions have been made in the literature~\cite{Li:2024psa,Ellis:2023oxs,Lozanov:2023rcd,Gelmini:2023kvo,Kitajima:2023cek,Geller:2023shn,Bai:2023cqj,Blasi:2023sej}.  
A more precise estimate based on the numerical simulations of the domain wall network suggests $\rho_{\rm DW} \simeq 0.5 \Delta V$~\cite{Saikawa:2017hiv,Hiramatsu:2010yz,Hiramatsu:2012sc,Hiramatsu:2013qaa}. 
}

\begin{figure}[t] 
\centering
 \includegraphics[width=0.55\linewidth]{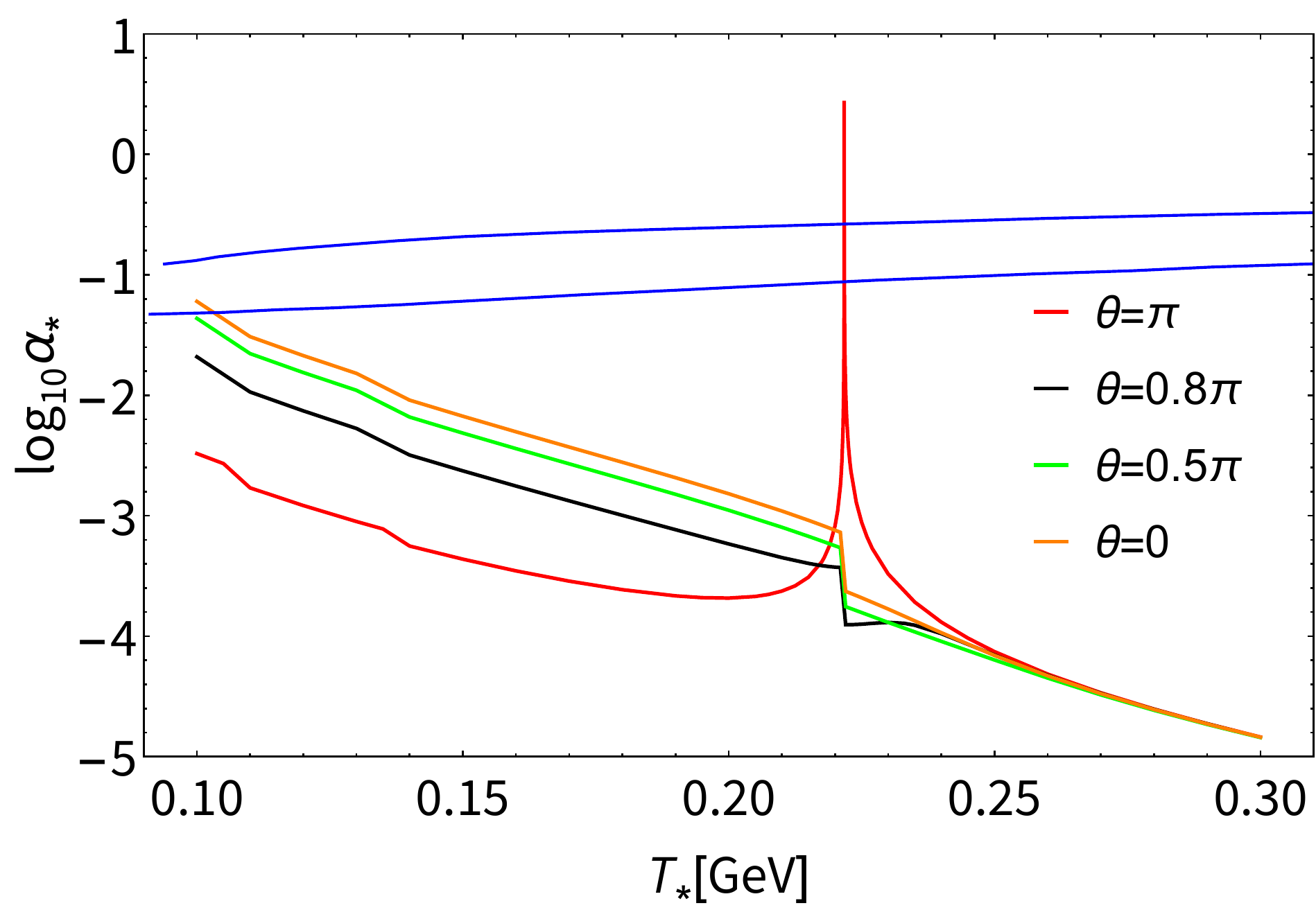}
 \caption{ The contour plot for the signal strength $\alpha_*(\theta)$ 
 in Eq.(\ref{alpha-star}) versus the ALP domain wall annihilation temperature $T_*$ with 
 $\chi_{\rm top}$ in Fig.~\ref{chitop-fig} encoded. 
 The 2$\sigma$ contour from Ref.~\cite{NANOGrav:2023hvm} 
 has also been displayed in blue,  
 and the regimes inside the 2$\sigma$ contour are thus allowed. 
 $\alpha_*$ in Eq.(\ref{alpha-star}) depends on the effective degrees of freedom $g_*(T)$ 
 around the QCD phase transition epoch, 
 which we have adjusted $g_*(T)$ available in Ref.~\cite{ParticleDataGroup:2022pth}, 
 in such a way that the deconfinement transition happens at the same time as the chiral phase transition 
 in the present NJL model takes place (at $T = 221$ MeV). 
 The wiggles seen around a lower $T$ regime 
 in the model prediction curves have thus been arisen from the thresholds encoded in $g_*(T)$. 
 } \label{alpha-T-contour-NJL}
\end{figure}

\textcolor{black}{
The QCD bias $|\chi_{\rm top}|$ generically does not significantly affect the ALP potential when the universe is hotter than the QCD scale of ${\cal O}(100)$ MeV, as seen from Fig.~\ref{chitop-fig} 
and also from the generic formula in Eq.(\ref{chitop}). 
This is essentially due to the correlation of the effective restoration of the chiral $SU(2)$ (via $\langle \bar{q}q\rangle$) and/or $U(1)$ axial symmetry ($\chi_\eta$) 
at higher temperatures (See Eq.(\ref{chitop})). 
We assume that the ALP-domain wall annihilation takes place at $T=T_*$ 
by the QCD bias and 
fully provides the source of the GW~\footnote{
\textcolor{black}{The ALP domain wall network will completely be decayed never to be left before the expected domain wall-dominated epoch arises. The critical temperature for the domain wall domination is estimated by using the present NJL model and evaluating the condition $\rho_{\rm DW}/\rho_{\rm rad} > 1$, to give $T= T_{\rm dom} \sim 19$ MeV, which is indeed much lower than the QCD scale.} 
}. 
The produced GW power spectrum at the peak frequency is then assessed via the signal strength,  
free from the domain wall string tension $\sigma_{\rm DW} \sim m_a f_a^2$, 
by the following signal strength (see, e.g., \cite{Blasi:2023sej})}:  
\begin{align} 
 \alpha_*(T_*, \theta) 
 &= 
 \frac{\rho_{\rm DW}(T_*)}{\rho_{\rm rad}(T_*)}\simeq  
 \frac{0.5 \Delta V(T_*, \theta)}{\rho_{\rm rad}(T_*)} 
 \notag\\ 
 &\simeq 0.15 
 \times 
 \left( \frac{|\chi_{\rm top}(T_*, \theta )|^{1/4}}{100\,{\rm MeV}} \right)^4 \left( \frac{T_*}{ 100\,{\rm MeV}} \right)^{-4}
 \left( \frac{g_*(T_*)}{10} \right)^{-1}
 \,,  \label{alpha-star}
\end{align}
where we have used $\rho_{\rm DW} \simeq 0.5 \Delta V$~\cite{Saikawa:2017hiv,Hiramatsu:2010yz,Hiramatsu:2012sc,Hiramatsu:2013qaa} and 
$\rho_{\rm rad}(T) = (\pi^2/30) g_*(T) T^4$. 
This $\alpha_*$ has been constrained by the NANOGrav 15yr data set as a function of $T_*$.  
See Fig.~\ref{alpha-T-contour-NJL}, showing that 
the cases with $ \theta/\pi \gtrsim 0.5$ are incompatible with the interpretation of the 
NANOGrav 15-year data at around $T_* = 100$ MeV. 
This is essentially due to the suppression of the scalar quark condensate 
as observed in $\chi_{\rm top}$ as shown in Fig.~\ref{chitop-fig}. 
At $\theta = \pi$ onset the criticality (at $T\simeq 221$ MeV), 
the signal gets dramatically enhanced by the singular spike structure 
reflecting the second order phase transition of the CP symmetry, 
which is, however, to be too big to interpret the data. 
This feature is insensitive to the deconfinement phase transition of QCD, which has not yet 
been incorporated in the figure.

\section{Summary and discussions} 

In summary, 
the ALP domain wall with 
a QCD bias may be incompatible with the NANOGrav 15-year data on a stochastic GW  background, when 
the domain wall network collapses in the hot-QCD induced local-CP odd domain 
which could allow to have a sizable QCD $\theta$ parameter. 
This is due to the drastic suppression of the QCD bias set by 
the QCD topological susceptibility, $\chi_{\rm top}$, in the presence of the CP-odd domain 
with $\theta = {\cal O}(1)$ (see Eq.(\ref{chi-top-suppress}) and Fig.~\ref{chitop-fig}). 
An explicit model analysis of $\chi_{\rm top}$ with a large $\theta$ 
based on the two-flavor NJL model implies 
that only at $\theta=\pi$, the QCD bias tends to get 
significantly large enough due to the criticality of the thermal CP restoration. 
However, this turned out to give too big signal strengths to be consistent with 
the NANOGrav 15-year data. 
\\

In closing, we give several comments on the issues to be pursed in the future.

\begin{itemize} 
\item 
The presently employed MFA of NJL model does not precisely 
 reproduce the chiral crossover at $\theta=0$ as has been observed 
 in the lattice simulations. The latter predicts the pseudocritical temperature 
 $T_{\rm pc} \simeq 155$ MeV~\cite{Aoki:2009sc,Borsanyi:2011bn,Ding:2015ona,Bazavov:2018mes,Ding:2020rtq}, while the present model yields $T_{\rm pc} \simeq 220$ MeV. 
This may imply a simple shift of the $T$-distribution of $\chi_{\rm top}$ in Fig.~\ref{chitop-fig} 
and also $\alpha_*(\theta)$ toward lower $T$ with lower criticality, say, at $T \sim 160$ MeV. 
Even in that case, however, the trend of the dramatic suppression of $\chi_{\rm top}$ with $\theta={\cal O}(1)$ should keep manifest for below the shifted criticality ($\sim 160$ MeV) 
and the spike signal associated with the strong CP restoration at $\theta = \pi$ will be left at the criticality. 

\item 
In Appendix~\ref{NJL-PNJL} we have also taken into account a sort of the QCD deconfinement phase transition by extending the NJL model into the Polyakov-loop NJL (PNJL) model~\cite{Fukushima:2003fw} (for a recent review, see, e.g., \cite{Fukushima:2017csk}). 
It turns out that the high suppression of $\chi_{\rm top}$ with $\theta={\cal O}(1)$ 
is still manifest and the presence of a sharp spike in $\chi_{\rm top}$ at $\theta=\pi$ still persists, 
in accord with the earlier work~\cite{Sakai:2011gs}. 
This implies the insensitivity of the deconfinement-confinement transition for the main conclusion addressed in the main text.

\item 
The local CP-odd domain would generate not only a large $\theta$, but also 
the fluctuation of $\theta$ in the temporal direction, $\partial_t \theta(t)$, 
which is identified as the so-called chiral chemical potential (often denoted as $\mu_5$)~\cite{Kharzeev:2007tn,Kharzeev:2007jp,Fukushima:2008xe}. 
The $\mu_5$ contribution to the thermal chiral phase transition as well as 
$\chi_{\rm top}$ has been discussed in Ref.~\cite{Ruggieri:2020qtq} based on a two-flavor NJL model with $\theta=0$.  
From the reference we can see that a part of $\chi_{\rm top}$, which only includes 
the scalar condensate term $\langle \bar{q}q \rangle$ as in Eq.(\ref{chitop-LS}) (without the $\chi_\eta$ contribution in Eq.(\ref{chitop})), roughly gets enhanced by about a factor of $(3/2)$ at 
around $T = {\cal O}(100)$ MeV. 
With $\theta = {\cal O}(1)$, the scalar condensate $\langle \bar{q}q \rangle$ generically 
gets smaller due to the CP violation as in the generic argument (Sec.~\ref{Sec-II}), which 
still holds even in the presence of $\mu_5$ because $\mu_5 \sim \partial_t \theta(t)$ 
does not modify the mixture structure between $(\bar{q}q)$ and $(\bar{q} i \gamma_5 q)$ as  
in Eq.(\ref{mixing}). 
Furthermore, the presence of $\mu_5$ as well as nonzero $\theta$ 
does not change the form of all the anomalous chiral Ward-identities as clarified in 
Appendix~\ref{ACWTIs}. 
Hence the high suppression of $\chi_{\rm top}$ would still be seen even with $\mu_5$, 
so the conclusion as presently claimed would be intact. 
More precise discussions including the size of the spike at $\theta=\pi$ 
would be worth pursing elsewhere. 

\item 
The criticality associated with the thermal CP restoration at $\theta=\pi$ 
would be crucial to precisely check if the ALP domain wall annihilation 
can still be viable to account for the NANOGrav 15-year data. 
The related spike structure, following the order of the phase transition, the first order or the second order, would be subject to effective model approaches for QCD~\cite{Dashen:1970et,Witten:1980sp,Pisarski:1996ne,Creutz:2003xu,Mizher:2008hf,Boer:2008ct,Creutz:2009kx,Boomsma:2009eh,Sakai:2011gs,Chatterjee:2011yz,Sasaki:2011cj,Sasaki:2013ewa,Aoki:2014moa,Gaiotto:2017tne,Verbaarschot:2014upa,Gaiotto:2017yup,Bai:2023cqj}. 
The presently employed NJL-type model with the MFA tends to predict the 
second order~\cite{Boer:2008ct,Boomsma:2009eh,Sakai:2011gs}, while the linear-sigma model type with the MFA~\cite{Pisarski:1996ne,Creutz:2003xu,Mizher:2008hf,Boer:2008ct,Creutz:2009kx,Boomsma:2009eh,Bai:2023cqj} and the `t Hooft anomaly matching argument~\cite{Gaiotto:2017tne,Gaiotto:2017yup} supports the first order 
phase transition. 
Even going beyond the MFA and/or the RPA, including subleading order contributions in $1/N_c$ expansion, the presently claimed incompatibility would be crucial to deduce a more definite 
conclusion of the compatibility of the ALP domain wall interpretation with 
NANOGrav 15-year data. 
The functional renormalization group analysis makes it possible to clarify 
the case, which deserves to another publication. 
At any rate, the presently claimed incompatibility would still pin down 
one benchmark point in the full parameter space of the QCD-biased ALP domain wall 
collapse.

\item 
\textcolor{black}{
As seen from Figs.~\ref{scalar-pseudo-conden} and~\ref{PNJL-scalar-pseudo-conden} in Appendix~\ref{NJL-PNJL}, 
at $\theta=\pi$ the thermal CP restoration point coincides with the pseudocritical point (inflection point) 
for the chiral crossover. This is due to the present MFA, which involves the two intrinsic features at $\theta=\pi$: 
i) the chiral symmetry is not spontaneously broken via the scalar quark condensate $\langle \bar{q}q \rangle$, 
as long as the CP symmetry is broken; 
ii) no $U(1)$ axial violating loop corrections are generated at this approximation level. 
Those lead to no thermal development of $\langle \bar{q}q \rangle$ until the CP symmetry restores, hence $\langle \bar{q}q \rangle$ gets kicked down instantaneously, i.e., undergoes the pseudocritical point at the same timing as the CP restoration. 
We have clarified those points in Appendix~\ref{NJL-PNJL} (around Eq.(\ref{pi-no-chiralSSB})). 
Going beyond the MFA analysis, such as the functional renormalization group, 
would potentially generate a gap between two critical points for the (effective) chiral and CP symmetry restorations, which should come from the $U(1)$ axial anomaly effects in loops.  
The further investigation along this line could also make a complementary  
benchmark compared to the prediction from the `t Hooft-anomaly matching in pure Yang-Mills theories~\cite{Chen:2020syd},  
which suggests the CP restoration temperature is higher than or equal to the deconfinement phase transition one. 
}

\item 
\textcolor{black}{
When strange quark contributions are incorporated in the analysis, 
the thermal CP restoration 
at $\theta = \pi$ would be more involved. }
First of all, one notices that 
the CP phase contribution carried by 
the strange quark is highly suppressed by a factor of $m_{u,d}/m_s$,  
as reviewed in Appendix~\ref{ACWTIs} (around Eqs.(\ref{FSN}) - (\ref{FS-Z})). 
This observation comes from the robust flavor singlet nature of the $\theta$ dependence in QCD that requires the strange quark field to carry the $\theta$ phase with 
the form $e^{i \frac{m_{u,d}}{2m_s} \theta} \simeq 1$, i.e., almost free from $\theta$. 
Thus, as far as the CP violating effects are concerned, 
the three-flavor QCD is essentially decomposed into 
the $2 + 1$ (the lightest two quarks and the strange quark) structure and 
the CP order parameter is almost controlled by the lightest two-flavor 
sector, i.e., $\sum_{q=u,d} \langle \bar{q} i \gamma_5 q  \rangle$.

This is the characteristic flavor violation and irrespective to the presence of the QCD topological charge fluctuation, which is flavor universal, or equivalently in 
the NJL framework, the `t Hooft-Kobayashi-Maskawa determinant term~\cite{Kobayashi:1970ji,Kobayashi:1971qz,tHooft:1976rip,tHooft:1976snw} (as in Appendix~\ref{NJL-PNJL}).  
In the three-flavor NJL with the MFA, thus the chiral crossover for $\langle \bar{s}s \rangle$ would still persist even at $\theta=\pi$, 
simply because of presence of the finite strange quark mass, 
while $\langle \bar{q}q \rangle$ would be subject to the approximation for QCD, or the (P)NJL.   
In the MFA of the three-flavor NJL, 
$\langle \bar{q}q \rangle$ 
would be trapped to a constant value until the CP phase transition takes place at $T=T_c$, and then starts to drops when $T > T_c$ (with discontinuity in the $T$-derivative at $T=T_c$), as in the two-flavor case (Fig.~\ref{scalar-pseudo-conden} in Appendix~\ref{NJL-PNJL}).   
As in the case with $\theta=0$, 
$\langle \bar{s}s \rangle$ would drop down with $T$ 
more slowly than $\langle \bar{q}q \rangle$ simply due to the larger strange quark mass.  
Thus, in the framework of the MFA of the NJL, the CP phase transition in 
the three-flavor case would be essentially identical to the one in the lightest two-flavor case, hence  
it would still be of the second order keeping the critical spike structure 
of $\chi_{\rm top}$, hence $\alpha_*$ does as well, 
as seen from Fig.~\ref{alpha-T-contour-NJL}.

\textcolor{black}{However, as has been discussed in the recent literature~\cite{Fejos:2016hbp,Fejos:2023lvw,Fejos:2021yod} based on 
the functional renormalization group analysis on the three-flavor linear sigma model with $\theta=0$, 
the `t Hooft-Kobayashi-Maskawa determinant term beyond the MFA could be nonperturbatively enhanced at around $T=T_c$ even in the case of $\theta=\pi$. 
This implies that the criticality of the CP phase transition as well as the chiral phase transition 
might significantly be altered and the entry of the strange quark contributions beyond the MFA might be nontrivial, though the carried CP phase is intrinsically highly suppressed as aforementioned.    
Thus, the detailed analysis in the three-flavor NJL, both within and beyond the MFA,  is noteworthy to pursue in another publication.  
}
 
\end{itemize}

Finally, we comment on impacts of lattice QCD calculations to this observation. 
Most of current lattice QCD calculations at the physical point have been done using the Monte-Carlo method. For nonzero theta, simulations are suffered from infamous sign problem. Namely, we cannot calculate expectation values with $|\theta| \gg 0$. 
To avoid the problem in the realistic setup, we employ imaginary theta and analytic continue to the real theta to get expectation values for physical observables~\cite{DelDebbio:2002xa,DElia:2003zne,DelDebbio:2006sbu,Giusti:2007tu,Izubuchi:2007rmy,Vicari:2008jw,DElia:2013uaf,Hirasawa:2024vns}. 
These calculations for $\chi_{\rm top}$ have been done away from the physical point.
An interesting methods for nonzero theta is suggested~\cite{Kitano:2021jho}, but it is still hard to get physical results.
Digital and analog quantum simulations and calculations with nonzero theta using tensor networks have been performed for toy models but not for the realistic setup like four dimensional QCD at the physical point~\cite{Byrnes:2002gj,Funcke:2019zna,Kuramashi:2019cgs,Chakraborty:2020uhf,Honda:2021aum,Honda:2022hyu}. 

\section*{Acknowledgments} 

We are grateful to Taishi Katsuragawa and Wen Yin for useful comments.  
This work was supported in part by the National Science Foundation of China (NSFC) under Grant No.11747308, 11975108, 12047569, 
and the Seeds Funding of Jilin University (S.M.), 
and Toyama First Bank, Ltd (H.I.).  
The work by M.K. was supported by the Fundamental Research Funds for the Central Universities 
and partially by the National Natural Science Foundation of China (NSFC) Grant  No. 12235016, and the Strategic Priority Research Program of Chinese Academy of Sciences under Grant No. XDB34030000.
The work of A.T. was partially supported by JSPS  KAKENHI Grant Numbers 20K14479, 22K03539, 22H05112, and 22H05111, and MEXT as ``Program for Promoting Researches on the Supercomputer Fugaku'' (Simulation for basic science: approaching the new quantum era; Grant Number JPMXP1020230411, and 
%
Search for physics beyond the standard model using large-scale lattice QCD simulation and development of AI technology toward next-generation lattice QCD; Grant Number JPMXP1020230409). 

\appendix

\section{Generic properties of anomalous chiral Ward-Takahashi identities (ACWTIs) in QCD}
\label{ACWTIs}

\subsection{ACWTIs with external gauge fields and $\theta$} 

We start with the QCD Lagrangian with $N_f$ quarks including the external gauge fields, 
\begin{align}
{\cal L}_{\rm QCD} 
= 
\bar q_Li\gamma^\mu D_\mu q_L
+ 
\bar q_Ri\gamma^\mu D_\mu q_R
-\bar q_L {\bf m}_f q_R - \bar q_R {\bf m}_f q_L
+\frac{1}{2}\sum_{a=1}^8{\rm tr}\left[(G_{\mu\nu}^a T^a_c)^2\right]
+\theta\frac{g^2}{64\pi^2}\epsilon^{\mu\nu\rho\sigma} G_{\mu\nu}^aG_{\rho\sigma}^a\,,
\end{align}
where $q _{L(R)}$ denotes the left-(right-) handed quark fields belong to the fundamental representation of $SU(N_f)$; 
${\bf m}_f$ is the current quark mass for $f$-quark taking 
the diagonal form like ${\bf m}_f = {\rm diag}( m_u, m_d, \cdots  )$; 
$G_{\mu\nu}^a$ $(a=0,1,\cdots,8)$ are the field strengths of the gluon fields $G_\mu^a$; 
$T^a_c$ stand for the generators of the QCD color group $SU(3)_c$; 
$g$ denotes the QCD gauge coupling constant; 
$\theta = \theta(x)$ plays the role of the source for the topological operator 
$i g^2\epsilon^{\mu\nu\rho\sigma} G_{\mu\nu}^aG_{\rho\sigma}^a$;  
The external gauge fields
${\cal L}_\mu^A$ and ${\cal R}_\mu^A$ ($A=0,\cdots , N^2_f -1$) are introduced by gauging the global 
chiral $U(N_f)_L$ and $U(N_f)_R$ symmetry, which are embedded into 
the covariant derivatives as 
\begin{align}
 D_\mu q_L 
 &= 
 \left(  \partial_\mu-ig G_\mu^a T^a_c 
 -i{\cal L}_\mu^AT_f^A\right)q_L\,,\nonumber\\
 D_\mu q_R 
 &= 
 \left( \partial_\mu-ig G_\mu^a T^a_c 
 -i{\cal R}_\mu^A T_f^A\right) q_R\,,
\end{align}
with $T_f^A$ being the generators of $U(N_f)$ in the flavor space.

The vacuum energy of QCD is given by 
\begin{align}
V_{{\rm QCD}}(\theta)=-i \ln Z_{{\rm QCD} }\,,
\end{align}
where $Z_{{\rm QCD}}$ represents the generating functional of QCD in Minkowski spacetime,
\begin{align}
Z_{\rm QCD}
=
\int [ dq  d\bar q ][dG]
\exp\Biggl[
i\int d^4x
{\cal L}_{\rm QCD}
\Biggl]\,.
\end{align}

We define the topological susceptibility $\chi_{\rm top}$ including the $\theta$ dependence:  
\begin{eqnarray}
\chi_{\rm top}(\theta)&=&
-\int d^4x\frac{\delta^2 V_{{\rm QCD}}(\theta)}{\delta \theta (x) \delta \theta (0)} 
-i
\int d^4x \frac{\delta V_{\rm QCD} }{\delta \theta (x)}
\frac{\delta V_{\rm QCD} }{\delta \theta (0)}
\notag\\ 
&=&
-i\int d^4x \left\langle0 \Big|T
\left(\frac{g^2}{64\pi^2}\epsilon^{\mu\nu\rho\sigma} G_{\mu\nu}^aG_{\rho\sigma}^a\right) (x)
\left(\frac{g^2}{64\pi^2}\epsilon^{\mu\nu\rho\sigma} G_{\mu\nu}^bG_{\rho\sigma}^b\right) (0) \Big|0 \right\rangle_\theta 
\label{chitop_Q}
\,, 
\end{eqnarray} 
where we have introduced the subscript $\theta$ to explicitize the $\theta$ dependence on the vacuum.


The external fields, such as the electromagnetic field $A_\mu$, 
the baryon chemical potential $\mu_{\rm B}$, 
and 
the chiral chemical potential $\mu_5$, are embedded in the external gauge  fields as~\footnote{ The isospin chemical potential can also be incorporated, when $N_f=2$, 
into the covariant derivative as an additional external gauge field term proportional 
to $(\sigma^3/2)$. } 
\begin{eqnarray}
{\cal L}_\mu^A T_f^A&=& \left[ eQ_{\rm em} A_\mu  
+\frac{\mu_{\rm B}}{3} {\bm 1}_{N_f \times N_f} 
\right]-\left[\mu_5 {\bm 1}_{N_f \times N_f} \right]
,\nonumber\\
{\cal R}_\mu^aT_f^a&=&
\left[ eQ_{\rm em} A_\mu  
+\frac{\mu_{\rm B}}{3} {\bm 1}_{N_f \times N_f} 
\right]+\left[\mu_5{\bm 1}_{N_f \times N_f} \right]
\,,
\end{eqnarray}
where 
$e$ represents the electromagnetic coupling constant, and $Q_{\rm em}$ denotes the electric charge matrix for quarks 
$Q_{\rm em}={\rm diag}(Q_{\rm em}^u, Q_{\rm em}^d,\cdots)$. 
The chemical potentials have been introduced as constant fields. 
In QCD with the external gauge fields, the $U(1)_A$ symmetry is explicitly broken by the current quark mass term and the gluonic quantum anomaly, and also external electromagnetic field. 
This is reflected in 
the anomalous conservation law of 
the $U(1)$ axial current for each quark in the $N_f$-plet quark field $q$, labeled as 
$q_f$, 
\begin{eqnarray} 
\partial_\mu j^{(f)\mu}_{A}
&=&
2i \bar q^f m_f\gamma_5 q^f+\frac{g^2}{32\pi^2}\epsilon^{\mu\nu\rho\sigma}  G _{\mu\nu}^a  G^a_{\rho\sigma} +N_c\frac{e^2 [Q_{\rm em}^f]^2 }{32\pi^2}\epsilon^{\mu\nu\rho\sigma} F_{\mu\nu}F_{\rho\sigma}\,,
\end{eqnarray}
with the $U(1)$ axial current $j_A^{(f)\mu}=\bar q^f\gamma_5\gamma^\mu q^f$.
Note that the constant chemical potentials do not contribute to the anomalous conservation law.

Under the $U(1)_A$ rotation with the rotation angle $\alpha_A$,
the quark fields transform as 
\begin{eqnarray}
q_L^f&\to& \exp(-i\alpha_A^f/2)q_L^{\prime f}\,,\nonumber\\
q_R^f&\to& \exp(+i\alpha_A^f/2)q_R^{\prime f}\,.
\end{eqnarray}
Then the QCD generating functional gets shifted as  
\begin{eqnarray}
&& \int [dq^\prime d\bar q^\prime][dG]
\exp\Biggl[
i\int d^4x
\Biggl(
\bar q_L^\prime i\gamma^\mu D_\mu q_L^\prime
+
\bar q_R^\prime i\gamma^\mu D_\mu q_R^\prime
-\sum_f\left( \bar q_L^{\prime f} m_f e^{i\alpha_A^f} q_R^{\prime f} + \bar q^{\prime f}_R m_f e^{-i\alpha_A^f} q^{\prime f}_L 
\right) 
+\frac{1}{2}\sum_{a=1}^8{\rm tr}\left[(G_{\mu\nu}^a T^a_c)^2\right]\nonumber\\
&&
+\left(\theta-\sum_f \alpha_A^f\right)\frac{g^2}{64\pi^2}\epsilon^{\mu\nu\rho\sigma} G_{\mu\nu}^aG_{\rho\sigma}^a
-\left(\sum_f \alpha_A^f Q_f^2\right)
N_c\frac{e^2 }{64\pi^2}\epsilon^{\mu\nu\rho\sigma} F_{\mu\nu}F_{\rho\sigma}
\Biggl)
\Biggl]\,.
\end{eqnarray} 
To rotate the QCD $\theta$ term away, we take the following phase choice reflecting 
the flavor-singlet nature of the QCD vacuum~\cite{Kawaguchi:2020kdl,Kawaguchi:2020qvg,Cui:2021bqf,Cui:2022vsr},
\begin{eqnarray}
\alpha_A^f = \frac{\bar m}{m_f} \theta
\,, \label{FSN}
\end{eqnarray}
with 
\begin{eqnarray}
\bar m=\left( \sum_f \frac{1}{m_f} \right)^{-1}\,.
\end{eqnarray}
Then the QCD generating functional goes like 
\begin{eqnarray}
Z_{\rm QCD}
&=& 
\int [dq^\prime  d\bar q^\prime ][dG]
\exp\Biggl[
i\int d^4x
\Biggl(
\bar q_L^\prime i\gamma^\mu D_\mu q_L^\prime
+
\bar q_R^\prime i\gamma^\mu D_\mu q_R^\prime
+\frac{1}{2}\sum_{a=1}^8{\rm tr}\left[(G_{\mu\nu}^a T^a_c)^2\right]
\nonumber\\
&&
- \sum_f\left(
\bar q^{\prime f}_L m_f \exp\left({i \frac{\bar m}{m_f} \theta}\right) q^{\prime f}_R
+ \bar q^{\prime f}_R m_f \exp\left({-i  \frac{\bar m}{m_f} \theta}\right) q^{\prime f}_L
\right) 
\notag\\ 
& & 
-\left(\sum_f \frac{\bar m}{m_f} \theta Q_f^2\right)
N_c\frac{e^2 }{64\pi^2}\epsilon^{\mu\nu\rho\sigma} F_{\mu\nu}F_{\rho\sigma}
\Biggl)
\Biggl]. 
\label{FS-Z}
\end{eqnarray}  
From this generating functional,
the topological susceptibility is given as 
\begin{eqnarray}
\chi_{\rm top}(\theta)&=&
\chi_{\rm top}^{(0)}(\theta) +\chi_{\rm top}^{\rm (EM)}(\theta)
\,, \label{full-chitop}
\end{eqnarray}
where 
\begin{eqnarray}
\chi_{\rm top}^{(0)}(\theta)
&=&
-\bar m^2
\Biggl[
\sum_f \frac{1}{m_f}
\left\langle0\left|
\left(
\bar q^{\prime f} q^{\prime f}  \cos\frac{\theta}{2}
+\bar q^{\prime f} i\gamma_5 q^{\prime f} \sin \frac{\theta}{2}
\right)\right|0\right\rangle_\theta \nonumber\\
&&
+
i\int d^4x
\left \langle 0 \left| 
T
\sum_f
 \left(
\bar q^{\prime f} i\gamma_5 q^{\prime f}\cos\frac{\theta}{2}
-\bar q^{\prime f} q^{\prime f}  \sin\frac{\theta}{2}
\right)
(x)
\sum_{f^\prime}
 \left(
\bar q^{\prime f^\prime} i\gamma_5 q^{\prime f^\prime} \cos\frac{\theta}{2}
-\bar q^{\prime f^\prime}q^{\prime f^\prime}  \sin\frac{\theta}{2}
\right)
(0)
\right|0
\right\rangle_\theta
\Biggl],
\nonumber\\
\chi_{\rm top}^{\rm (EM)}(\theta)
&=&
-i
\left(\sum_f \frac{\bar m Q_f^2}{m_f}\right)^2
\int d^4x \left\langle0 \Big|T
\left(
\frac{e^2 N_c }{64\pi^2}\epsilon^{\mu\nu\rho\sigma} F_{\mu\nu}F_{\rho\sigma}
\right)(x)
\left(
\frac{e^2 N_c }{64\pi^2}\epsilon^{\mu\nu\rho\sigma} F_{\mu\nu}F_{\rho\sigma}
\right)(0)
\Big|0 \right\rangle_\theta   \nonumber\\
&&
-i
\left(\sum_f \frac{\bar m^2 Q_f^2}{m_f}\right)
\int d^4x \left\langle0 \Big|T
\left(
\frac{e^2 N_c }{64\pi^2}\epsilon^{\mu\nu\rho\sigma} F_{\mu\nu}F_{\rho\sigma}
\right)(x)
\sum_{f}
 \left(
\bar q^{\prime f} i\gamma_5 q^{\prime f} \cos\frac{\theta}{2}
-\bar q^{\prime f}q^{\prime f}  \sin\frac{\theta}{2}
\right)
(0)
\Big|0 \right\rangle_\theta\nonumber\\
&&
-i
\left(\sum_f \frac{\bar m^2 Q_f^2}{m_f}\right)
\int d^4x \left\langle0 \Big|T
 \left(
\bar q^{\prime f} i\gamma_5 q^{\prime f} \cos\frac{\theta}{2}
-\bar q^{\prime f}q^{\prime f}  \sin\frac{\theta}{2}
\right)
(x)
\left(
\frac{e^2 N_c }{64\pi^2}\epsilon^{\mu\nu\rho\sigma} F_{\mu\nu}F_{\rho\sigma}
\right)(0)
\Big|0 \right\rangle_\theta
\,. \label{chitop-EM}
\end{eqnarray} 
These susceptibilities are written in terms of the primed quark field $q^\prime$. They can be rewritten in terms of the original quark field $q$ using the following connection, 
\begin{align}
q= \exp(i\theta \gamma_5/4)q^\prime.
\end{align}
Then
$\chi_{\rm top}^{(0)}(\theta)$ and 
$\chi_{\rm top}^{\rm (EM)}(\theta)$ go like
\begin{align}
\chi_{\rm top}^{(0)}(\theta) 
&= 
-\bar m^2
\left[
\sum_f \frac{1}{m_f}\langle0| \bar q^f q^f|0\rangle
+i\int d^4x \left\langle0 \Big| T 
\left(\sum_f\bar q^f i\gamma_5 q^f\right)(x)
\left(\sum_{f^\prime}\bar q^{f^\prime} i\gamma_5 q^{f^\prime}\right)(0)
\Big|0 \right\rangle_\theta\right],\nonumber\\
\chi_{\rm top}^{\rm (EM)}(\theta) 
&= 
-i
\left(\sum_f \frac{\bar m Q_f^2}{m_f}\right)^2
\int d^4x \left\langle0 \Big|T
\left(
\frac{e^2 N_c }{64\pi^2}\epsilon^{\mu\nu\rho\sigma} F_{\mu\nu}F_{\rho\sigma}
\right)(x)
\left(
\frac{e^2 N_c }{64\pi^2}\epsilon^{\mu\nu\rho\sigma} F_{\mu\nu}F_{\rho\sigma}
\right)(0)
\Big|0 \right\rangle_\theta\nonumber\\
&
-i
\left(\sum_f \frac{\bar m^2 Q_f^2}{m_f}\right)
\int d^4x \left\langle0 \Big|T
\left(
\frac{e^2 N_c }{64\pi^2}\epsilon^{\mu\nu\rho\sigma} F_{\mu\nu}F_{\rho\sigma}
\right)(x)
\left(\sum_f\bar q^f i\gamma_5 q^f\right)(0)
\Big|0 \right\rangle_\theta\nonumber\\
& 
-i
\left(\sum_f \frac{\bar m^2 Q_f^2}{m_f}\right)
\int d^4x \left\langle0 \Big|T
\left(\sum_f\bar q^f i\gamma_5 q^f\right)(x)
\left(
\frac{e^2 N_c }{64\pi^2}\epsilon^{\mu\nu\rho\sigma} F_{\mu\nu}F_{\rho\sigma}
\right)(0)
\Big|0 \right\rangle_\theta
\,. \label{chitops}
\end{align} 
The $\chi_{\rm top}$ term in Eq.(\ref{full-chitop}) is precisely the generalization 
of the one in Eq.(\ref{chitop}) with Eq.(\ref{chi-eta}), which has also been addressed  
in the literature~\cite{Kawaguchi:2020kdl,Kawaguchi:2020qvg,Cui:2021bqf,Cui:2022vsr} without external gauge fields. 
Thus the presence of the electromagnetic axial anomaly yields additional 
topological susceptibility, $\chi_{\rm top}^{({\rm EM})}$.

%
%




Next, we evaluate the ACWTIs for $SU(N_f)_L \times 
SU(N_f)_R$ transformation in QCD. 
Under the chiral $SU(N_f)$ transformation, the quark field $q$ transforms as
\begin{eqnarray}
q(x)&\to& 
q(x)+ 
i\alpha^A(x) T^A \gamma_5 q(x)\,, 
\qquad \, A= 1, \cdots N_f^2 - 1 
\,. 
\end{eqnarray}
 Then, the chiral $SU(N_f)$ transformation of
 the expectation value for an arbitrary local operator ${\cal O}(x_1)$ in the path integral 
formalism yields the ACWTIs: 
\begin{eqnarray}
\int d^4x
\lim_{\alpha^A \to0}
\left\langle 
\frac{ \delta{\cal O}'(x_1)}{\delta \alpha^A(x)}
\right\rangle_\theta 
+
\int d^4x
\left\langle 
{\cal O}(x_1) i
D_\mu j_5^{A \mu}(x)
\right\rangle_\theta 
+
\int d^4x
\left\langle 
{\cal O}(x_1) 
\bar q\left\{{\bf m}_f, T^A
\right\}\gamma_5q(x)
\right\rangle_\theta =0
\label{AWI_O-1}
\,, 
\end{eqnarray}
where 
$j_5^{A \mu}$ denotes the chiral $SU(N_f)$ current,
$j_5^{A \mu}=\bar q \gamma^\mu \gamma_5  T^A q$, and 
\begin{eqnarray}
D_\mu j_5^{A \mu}&=&
\partial_\mu 
j_5^{A \mu} - i[ {\cal V}_\mu, j_5^{A \mu}] 
\,, 
\end{eqnarray}
with ${\cal V}_\mu = ({\cal R}_\mu + {\cal L}_\mu)/2$. 
Equation (\ref{AWI_O-1}) is a generalization of the ACWTIs 
addressed in 
~\cite{Kawaguchi:2020kdl,Kawaguchi:2020qvg,Cui:2021bqf,Cui:2022vsr} 
without external gauge fields. 
Unless external gauge fields possess a topologically nontrivial configuration, we can rewrite the covariant derivative term as  
\begin{eqnarray}
\int d^4x\left\langle {\cal O}(x_1) iD_\mu j_5^{A \mu}(x)\right\rangle_\theta 
=
i\int d^4x
D_\mu^{(x)}
\left\langle 0|T
{\cal O}(x_1) 
 j_5^{A \mu}(x)
|0\right\rangle_\theta
=  \textrm{surface term} 
\,. 
\end{eqnarray} 
Thus we eventually have 
\begin{eqnarray}
\int d^4x
\lim_{\alpha^A \to0}
\left\langle 
\frac{ \delta{\cal O}'(x_1)}{\delta \alpha^A (x)}
\right\rangle_\theta 
+
\int d^4x
\left\langle 
{\cal O}(x_1) 
\bar q\left\{{\bf m}_f, T^A 
\right\}\gamma_5 q(x)
\right\rangle_\theta =0
\label{AWI_O}
\,. 
\end{eqnarray} 
This implies that the ACWTIs keep the same form as those in the case without 
external gauge fields~\cite{Kawaguchi:2020kdl,Kawaguchi:2020qvg,Cui:2021bqf,Cui:2022vsr}.

For instance, in the case of $N_f=3$, 
we find the same form of the ACWTIs as in the literature~\cite{GomezNicola:2016ssy,Kawaguchi:2020qvg,Cui:2021bqf,Cui:2022vsr}:   
\begin{align} 
\langle \bar uu \rangle_\theta +\langle \bar dd \rangle_\theta 
 &= - m_l \chi_\pi(\theta)  
\,, \notag\\  
\langle \bar uu \rangle_\theta +\langle \bar dd \rangle_\theta
+ 4 \langle \bar s s \rangle_\theta
& = 
- \left[ m_l 
\left(\chi_{P}^{uu}+\chi_{P}^{dd}+2\chi_{P}^{ud}\right)
- 2 (m_s + m_l)\left(\chi_{P}^{us}+\chi_{P}^{ds} \right)
+ 4 m_s \chi_P^{ss} \right]_\theta  
\,, \notag\\ 
\langle \bar uu \rangle_\theta +\langle \bar dd \rangle_\theta
-2  \langle \bar s s \rangle_\theta
& = 
- \left[ m_l 
\left(\chi_{P}^{uu}+\chi_{P}^{dd}+2\chi_{P}^{ud}\right)
+  (m_l - 2 m_s)\left(\chi_{P}^{us}+\chi_{P}^{ds} \right) 
-  2 m_s \chi_P^{ss} \right]_\theta 
		\,,  \label{AWTIs:Nf3}
	\end{align} 
with $m_l = m_u = m_d$.  
Here $\chi_{\pi}(\theta)$ denotes the pion susceptibility defined as 
\begin{align} 
	\chi_{\pi}(\theta)
	&= \int_T  d^4 x 
	\left[ 
	\langle ( \bar u(0) i \gamma_5  u(0))( \bar u(x) i\gamma_5 u(x))\rangle_{\rm conn}
+ \langle ( \bar d(0)  i\gamma_5 d(0))( \bar d(x) i\gamma_5 d(x))\rangle_{\rm conn}
\right]_\theta 
\,,  		\label{chipi-def}
	\end{align} 
with 
$\langle \cdot \cdot \cdot \rangle_{\rm conn}$ being the connected part of the correlation function, 
and the pseudoscalar susceptibilities  
$\chi_P^{uu,dd,ud}$, $\chi_P^{ss}$ and $\chi_P^{us, ds}$ 
are defined as 
\begin{align} 
        \chi_P^{f_1f_2}(\theta) &=\int_T  d^4 x \langle (\bar q_{f_1}(0) i \gamma _5 q_{f_1}(0))(\bar q_{f_2}(x) i \gamma _5 q_{f_2}(x))\rangle_\theta
\,, \qquad {\rm for} \quad {q_{_{f_{1,2}}} = u,d,s} 
\,.  		\label{psesus}
	\end{align} 
$\chi_{\rm top}^{(0)}(\theta)$ in Eq.(\ref{chitops}) then takes a couple of equivalent  forms, related each other by the ACWTIs in Eq.(\ref{AWTIs:Nf3})~\cite{Kawaguchi:2020kdl,Kawaguchi:2020qvg,Cui:2021bqf,Cui:2022vsr}
	\begin{align}		
 \chi_{\rm top}^{(0)}(\theta ) 
		&=
		 \bar{m}^2 \left[ 
		\frac{\langle \bar {u}u \rangle}{m_l}  
		+\frac{\langle \bar {d}d \rangle}{m_l} 
		+\frac{\langle \bar {s}s \rangle}{m_s} 
		+  
		\chi_{P}^{uu}+\chi_{P}^{dd}
		+\chi_P^{ss} 
		+ 2 \chi_P^{ud} 
		+2\chi_{P}^{us}+
		2\chi_{P}^{ds}
        \right]_\theta 
	\notag \\ 
	& = \frac{1}{4} \left[
	m_l\left(\langle \bar{u} u \rangle +\langle \bar{d} d \rangle  \right)
	+ m_l^2\left(\chi_{P}^{uu}+\chi_{P}^{dd}
	+2\chi_{P}^{ud}
	\right)
	\right]_\theta 
	\notag \\ 
&= \frac{m_l^2}{4}\left(\chi_\pi(\theta)-\chi_\eta(\theta) \right) 
\notag\\ 
 & = m_s \langle \bar{s}s \rangle_\theta + m_s^2 \chi_P^{ss}(\theta) 
	\,, \label{chitop:Nf3}
	\end{align}
with $\bar{m} = \left( \frac{2}{m_l} + \frac{1}{m_s} \right)^{-1}$. 
It is interesting to note that $\chi_\pi$ and $\chi_\eta$ are 
related not to the full topological susceptibility $\chi_{\rm top}$, but 
$\chi_{\rm top}^{(0)}$, in Eq.(\ref{full-chitop}).

\subsection{Renormalization group invariance of $\chi_{\rm top}$}

The current quark mass parameter is multiplicatively 
renormalized as 
\begin{eqnarray}
m_f^{0}(\Lambda)=Z_S^{-1}(\Lambda,\mu) m_f^R(\mu)
\,, 
\end{eqnarray}
where $\Lambda$ denotes the bare cutoff and the upper script $R$ 
stands for the quantity renormalized at the scale $\mu$. 
When the mass independent renormalization is thus applied. 
the renormalization factor $Z_S^{-1}$ is flavor universal.
Then we are allowed to drop the flavor label $f$ as 
\begin{eqnarray}
\bar m^0(\Lambda) =Z_S^{-1}(\Lambda,\mu)  
\bar m^R(\mu).
\end{eqnarray}
This also implies that 
$Z_S$ is independent of the current quark mass as well.

The $\bar q_fq_f$ bilinear operator is also multiplicatively 
renormalized by $Z_S(\lambda,\mu)$, 
\begin{eqnarray}
\left(\bar q_fq_f\right)_R=
Z_S^{-1}(\Lambda,\mu)
\left(\bar q_fq_f\right)_\Lambda
\,, 
\end{eqnarray}
so that we have the renormalization group invariant mass term like 
\begin{eqnarray}
m_f^{0}(\Lambda)\left(\bar q_fq_f\right)_\Lambda
=m_f^R(\mu) \left(\bar q_fq_f\right)_R 
\,. 
\end{eqnarray}

Consider also a ($\bar q_{f}i\gamma _5 q_{f}$) operator to be  renormalized in a similar way with the renormalization constant $Z_P^{-1}(\lambda,\mu)$
\begin{eqnarray}
(\bar q_{f}i\gamma _5 q_{f})_R=
Z_P^{-1}(\Lambda,\mu)
(\bar q_{f}i\gamma _5 q_{f})_\Lambda.
\end{eqnarray}
Since none of quark mass dependence is generated in the mass independent renormalization, 
the $U(1)$ axial invariance keeps manifest between renormalization of the ($\bar{q}_f q_f$)  and ($\bar{q}_f i \gamma_5 q_f$) operators. 
Hence we have 
\begin{eqnarray} 
Z_P^{-1}= Z_S^{-1}\,. 
\end{eqnarray} 
In that case we also find  
\begin{eqnarray}
\left(\chi_P^{f_1f_2}\right)_\Lambda=
Z_S^2
\left(\chi_P^{f_1f_2}\right)_R
\,, 
\end{eqnarray}

Now, we apply the renormalization procedure as above to 
$\chi_{\rm top}^{(0)}$ in the three-flavor case (Eq.(\ref{chitop:Nf3})): 
\begin{eqnarray}
\left(\chi_{\rm top}^{(0)} \right)_{\Lambda}
&=&
 -\bar{m}_0^2 \left[ 
\frac{\langle (\bar {u}u)_\Lambda \rangle}{m_u^0}  
		+\frac{\langle (\bar {d}d)_\Lambda \rangle}{m_d^0} 
		+\frac{\langle (\bar {s}s)_\Lambda \rangle}{m_s^0} 
		+  
		i(\chi_{P}^{uu})_\Lambda+i(\chi_{P}^{dd})_\Lambda
		+i(\chi_P^{ss} )_\Lambda
		+ 2i(\chi_P^{ud} )_\Lambda
		+2i(\chi_{P}^{us})_\Lambda+
		2i(\chi_{P}^{ds} )_\Lambda
        \right]\nonumber\\
&=&
-\bar{m}_R^2 \left[ 
\frac{\langle (\bar {u}u)_R \rangle}{m_u^R}  
		+\frac{\langle (\bar {d}d)_R \rangle}{m_d^R} 
		+\frac{\langle (\bar {s}s)_R \rangle}{m_s^R} 
		+  
		i(\chi_{P}^{uu})_R+i(\chi_{P}^{dd})_R
		+i(\chi_P^{ss} )_R
		+ 2i(\chi_P^{ud} )_R
		+2i(\chi_{P}^{us})_R+
		2i(\chi_{P}^{ds} )_R
        \right]\nonumber\\
&=&\left(\chi_{\rm top}^{(0)} \right)_{R}\,. 
	\end{eqnarray}
Thus it has been proven that $\chi_{\rm top}^{(0)}$ is renormalization group invariant. 
Note that this argument is also applicable to the case with nonzero $\theta$, 
because the QCD $\theta$ is itself conventionally renormalization group invariant 
and no new divergent terms induced due to nonzero $\theta$ will be generated, hence 
the renormalization factors will not be corrected. 
Furthermore, one can readily see that the external gauge-induced $\chi_{\rm top}^{\rm EM}$, defined as in Eq.(\ref{chitop-EM}), is also manifestly renormalization group invariant.

\section{The details on the (P) NJL model analysis} 
\label{NJL-PNJL}

\subsection{NJL case}

Our reference NJL Lagrangian with two flavors (up and down quarks)  and 
nonzero $\theta$ follows the literature~\cite{Sakai:2011gs}, which takes the form  
\begin{align} 
{\cal L} = \bar{q} (i \gamma_\mu \partial^\mu - m) q 
+ \frac{g_s}{2} \sum_{a=0}^3 \left[ (\bar{q} \tau_a q)^2 + (\bar{q} i \gamma_5 \tau_a q)^2 \right] 
+ g_d \left( e^{i\theta} \mathop{\rm det} [\bar{q} (1+ \gamma_5) q] 
+  e^{-i\theta} \mathop{\rm det}  [\bar{q} (1- \gamma_5) q] 
\right)
\,, \label{NJL-Lag}
\end{align}
where $q=(u, d)^T$ and the determinant acts on the quark flavors.  
The $g_d$ term, called the `t Hooft-Kobayashi-Maskawa determinant term~\cite{Kobayashi:1970ji,Kobayashi:1971qz,tHooft:1976rip,tHooft:1976snw}, breaks $U(1)$ axial symmetry, but keeps $SU(2)_L \times SU(2)_R \times U(1)_V$ symmetry, 
while others keep the full chiral $U(2)_L \times U(2)_R$ symmetry. 
The $g_d$ term thus serves as the $U(1)$ axial anomaly: 
\begin{align} 
\partial_\mu j^\mu_A=2i \bar q m \gamma_5 q
- 8 g_d {\rm Im} \left[\mathop{\rm det} \bar q(1-\gamma_5)q \cdot e^{-i\theta} \right].
\end{align}

Since the $U(1)$ axial symmetry is broken only by $g_d$ and the quark mass terms, 
one can move $\theta$ in the $g_d$ term to the mass term by a $U(1)$ axial rotation 
as was done in the case of the general argument above:  
\begin{align} 
 q \to e^{-i \gamma_5 \frac{\theta}{2}} q \equiv q' 
 \,, \label{prime-trans}
\end{align}
so that 
\begin{align} 
{\cal L} \to {\cal L}' = \bar{q}' (i \gamma_\mu \partial^\mu - m(\theta) ) q' 
+ \frac{g_s}{2} \sum_{a=0}^3 \left[ (\bar{q}' \tau_a q')^2 + (\bar{q}' i \gamma_5 \tau_a q')^2 \right] 
+ g_d \left( \mathop{\rm det} [\bar{q}' (1+ \gamma_5) q'] 
+ {\rm h.c.}  
\right)
\,,  
\end{align}
where 
\begin{align} 
m(\theta) =  m \left[ \cos \frac{\theta}{2} + i \gamma_5 \sin\frac{\theta}{2} \right]
\,,  
\end{align}
in which nonzero $\theta$ manifestly signals the CP violation. 
Use of this "prime" basis is convenient to analyze the model 
because all the $\theta$ dependence is 
transformed and collected into the complex mass $m(\theta)$ in the quark propagator. 
Similarly to Eq.(\ref{q-mass-term}), 
Eq.(\ref{prime-trans}) relates 
the scalar and pseudoscalar bilinears between the original- and prime-base scalar 
and pseudoscalar bilinears (for each quark flavor $i$) as 
\begin{align} 
 (\bar{q}_iq_i) &= (\bar{q}_i' q_i') \cos\frac{\theta}{2} + (\bar{q}_i' i \gamma_5 q_i') \sin\frac{\theta}{2}   
 \,, \notag\\ 
 (\bar{q}_i i \gamma_5 q_i) &= - (\bar{q}_i' q_i') \sin\frac{\theta}{2} + (\bar{q}_i' i \gamma_5 q_i') \cos\frac{\theta}{2}   
\,. \label{rot}
\end{align}

We work in the MFA, so that 
the scalar and pseudoscalar bilinears $(\bar{q}_i' q_i')$ and $(\bar{q}_i' i \gamma_5 q_i')$ 
are expanded around the means fields $S' = \langle \bar{q}_i' q_i' \rangle$  
and $P' = \langle \bar{q}_i'i \gamma_5 q_i' \rangle$, 
as $\bar{q}_i' q_i'  = S' + (: \bar{q}_i' q_i'  :)$ and 
$\bar{q}_i' i \gamma_5 q_i'  = P' + (: \bar{q}_i' i \gamma_5 q_i'  :)$, where 
the terms sandwiched by ``$:$" stand for the 
normal ordered product, meaning that $\langle : {\cal O} :  \rangle  =0$ 
for ${\cal O} = S', P'$. 
Then 
the interaction terms in Eq.(\ref{NJL-Lag}) are replaced, up to the normal ordered terms, as 
\begin{align} 
(\bar{q}_i' q_i')^2 & \to  4 S' \sum_i (\bar{q}_i' q_i') - 4 S'^2   
\,, \notag \\ 
(\bar{q}_i' i \gamma_5  q_i')^2 & \to 4 P' \sum_i (\bar{q}_i' i \gamma_5  q_i') - 4 P'^2 
\, , \notag\\ 
{\rm det}[\bar{q}_i' (1 + \gamma_5) q_i'] + {\rm h.c.} &\to 
2 S' \sum_i (\bar{q}_i' q_i') - 2 P'\sum_i (\bar{q}_i' i \gamma_5  q_i') - 2 (S'^2 - P'^2) 
\,. 
\end{align}
In the MFA 
the NJL Lagrangian in Eq.(\ref{NJL-Lag}) thus takes the form 
\begin{align} 
{\cal L}_{\rm MFA} = \sum_i \bar{q}'_i (i \gamma_\mu \partial^\mu - {\cal M}(S', P'; \theta)) q_i' 
- 2 g_s (S'^2 + P'^2) - 2g_d (S'^2 - P'^2)
\,,  
\end{align}
with 
\begin{align} 
{\cal M}(S', P'; \theta) 
&= \alpha(S'; \theta) + i \gamma_5 \beta(P'; \theta) 
\,, \notag\\ 
\alpha(S' ; \theta) &= 
m \cos\frac{\theta}{2} - 2 (g_s + g_d)S' 
\,, \notag\\ 
\beta( P'; \theta) 
&= 
m \sin\frac{\theta}{2} - 2 (g_s - g_d)P'
\,.  \label{alpha-beta}
\end{align}

Integrating out quarks leads to the thermodynamic potential in the MFA: 
\begin{align} 
 \Omega[S', P'; \theta] 
 &= 2 g_s (S'^2 + P'^2) + 2g_d (S'^2 - P'^2) 
 - 2 N_c N_f \int \frac{d^3 k}{(2 \pi)^3} \left[ E + 2 T \ln(1+ e^{-E/T})\right]
\,, \label{Omega}
\end{align} 
where $N_f=2$ and $N_c=3$, and 
\begin{align} 
E = \sqrt{M^2 + k^2} \,, \qquad 
M^2 
= \alpha^2 + \beta^2 \,, 
\label{M}
\end{align}
with $k^2 = |\vec{k}|^2$. 
Then $\langle \bar{q}_i' q_i' \rangle = S'$ 
and $\langle \bar{q}_i'i \gamma_5 q_i' \rangle = P'$ are determined  
through the stationary condition, 
\begin{align} 
 \frac{\partial \Omega}{\partial S'} = \frac{\partial \Omega}{\partial P^\prime} = 0 
\label{stationary} \,. 
\end{align}

\textcolor{black}{Of particular interest is to see the thermodynamic potential at $\theta=\pi$, 
\begin{eqnarray}
 \Omega[\alpha, \beta; \theta=\pi] &=& 
\frac{1}{2g_s+2g_D}\alpha^2
+
\frac{1}{2g_s-2g_D}
(\beta-m)^2
- 2 N_c N_f \int \frac{d^3p}{(2\pi)^3} \left[
E + 2 T \ln \left( 1+ e^{-E/T}      \right)
\right]\,, 
\label{Omega-pi}
\end{eqnarray}
where things have been written in terms of $\alpha$ and $\beta$ by means of Eq.(\ref{alpha-beta}). In the presently applied MFA, the loop corrections (corresponding to the last term in Eq.(\ref{Omega-pi})) are $U(1)$ axial invariant, i.e., do not separate $\alpha$ and $\beta$ due to no loop contributions involving vertices with the determinant coupling $g_d$. 
At $\theta=\pi$, the explicit-chiral $SU(2)$ breaking-effect (by $m$) has completely been transported into the $\beta$ direction (the second term).  
The stationary condition then takes the form   
\begin{eqnarray}
\alpha &=&
2 N_c N_f 
(g_s+g_D)\alpha
\int \frac{d^3p}{(2\pi)^3} 
\frac{1}{E}
\left[
1 -\frac{2}{  1+ e^{E/T}      }
\right],\nonumber\\ 
\beta-m&=&
2 N_c N_f 
(g_s-g_D)\beta
\int \frac{d^3p}{(2\pi)^3} 
\frac{1}{E}
\left[
1 -\frac{2}{  1+ e^{E/T}      }
\right]\,. 
\end{eqnarray}
When $\alpha\neq 0$, i.e., the CP symmetry is spontaneously broken, 
the gap equation for $\beta$ can be rewritten by eliminating the loop correction part 
as 
\begin{eqnarray}
&&\beta-m = \frac{g_s-g_D}{g_s+g_D}\beta\nonumber\\
&\leftrightarrow&
\beta = \frac{g_s+g_D}{2g_D}m\nonumber\\
&\leftrightarrow&
P^\prime = -\frac{m}{4g_D}
\,. \label{pi-no-chiralSSB}
\end{eqnarray}
This implies that in the CP broken phase at $\theta=\pi$, 
the chiral $SU(2)$ symmetry is not spontaneously broken and the chiral order 
parameter does not evolve in $T$. }


We come back to the case with arbitrary $\theta$ and derive relevant formulae for 
the susceptibilities $\chi_\eta$ and $\chi_{\rm top}$ within the MFA. 
First, by noting the change of basis in Eq.(\ref{rot}), 
the $\eta$ susceptibility $\chi_\eta$ in $\chi_{\rm top}$ of Eq.(\ref{chitop}) is 
written in terms of the prime base as 
\begin{align} 
\chi_\eta = \cos^2 \frac{\theta}{2} \chi^\prime_\eta - 2 \sin \frac{\theta}{2} \cos \frac{\theta}{2} \chi_{\eta \sigma}^\prime +  \sin^2 \frac{\theta}{2} \chi^\prime_\sigma 
\,,  
\end{align} 
The ``primed" meson susceptibilities are evaluated in the so-called random phase approximation 
(the bubble-ring resummation ansatz) as 
\begin{align} 
{\bf X}' &= {\bf \Pi}' \cdot \frac{1}{{\bf 1}_{2 \times 2} + {\bf G}' \cdot {\bf \Pi}'} 
\,, \label{X-prime}
\end{align}
where 
\begin{align} 
 {\bf X}' & = 
 \left( 
 \begin{array}{cc} 
\chi_\sigma' & \chi_{\eta \sigma}' \\ 
\chi_{\eta \sigma}' & \chi_\eta' 
\end{array} 
 \right) \,, \notag \\ 
  {\bf \Pi}' & = 
 \left( 
 \begin{array}{cc} 
\Pi_\sigma' & \Pi_{\eta \sigma}' \\ 
\Pi_{\eta \sigma}' & \Pi_\eta' 
\end{array} 
 \right) \,, \notag \\ 
 {\bf G}' & = 
 \left( 
 \begin{array}{cc} 
G_\sigma' & G_{\eta \sigma}' \\ 
G_{\eta \sigma}' & G_\eta' 
\end{array} 
 \right) 
=  
 \left( 
 \begin{array}{cc} 
g_s + g_d & 0  \\ 
0 & g_s - g_d  
\end{array} 
 \right) 
 \,,   
\end{align}
with the vacuum polarization functions for each channel, 
\begin{align} 
 \Pi_\sigma' & = 2 I_{S'} 
 \,, \notag\\ 
 \Pi_{\eta \sigma} &= I_{S'P'} 
 \,, \notag\\ 
 \Pi_{\eta}'&= 2 I_{P'} 
 \,, \notag\\ 
 I_{S'} & = - \frac{N_c}{\pi^2} \int_0^\Lambda d k k^2 \frac{E^2 - \alpha^2}{E^3} \left[1 - \frac{2}{e^{E/T} + 1}  \right] 
 \,, \notag\\ 
  I_{S'P'} & =  \frac{N_c}{\pi^2} \int_0^\Lambda d k k^2 \frac{2 \alpha \beta}{E^3} \left[1 - \frac{2}{e^{E/T} + 1}  \right] 
 \,. \notag\\ 
   I_{P'} & = - \frac{N_c}{\pi^2} \int_0^\Lambda d k k^2 \frac{E^2 - \beta^2}{E^3} \left[1 - \frac{2}{e^{E/T} + 1}  \right] 
 \,.  \label{Is}
\end{align} 
In evaluating the vacuum polarization functions, we have 
regularized the 3-momentum integrals by the cutoff $\Lambda$.

Putting all the relevant things above 
into $\chi_{\rm top}$ in Eq.(\ref{chitop}), we compute 
$\chi_{\rm top}$ as a function of $T$ and $\theta$. 
The model parameter setting follows the literature~\cite{Boer:2008ct,Boomsma:2009eh,Sakai:2011gs}: 
\begin{align} 
 m &= 6 \, {\rm MeV} 
\,, \notag\\ 
\Lambda & = 590\,{\rm MeV} 
\,, \notag\\ 
 g_s &=  2 (1 - c) G_0 \,
 \,, \qquad  
 g_d = 2 c G_0 \, \qquad 
 {\rm with} \qquad G_0 \Lambda^2 = 2.435 \qquad {\rm and} \qquad c=0.2
\,. \label{SC-NJL}
\end{align} 
\textcolor{black}{The $g_d$ coupling has been related by $c$ to the $g_s$ coupling just in a numerical manner, though the associated asymmetry features are different. }


\begin{figure}[t] 
\centering
	\includegraphics[width=0.46\linewidth]{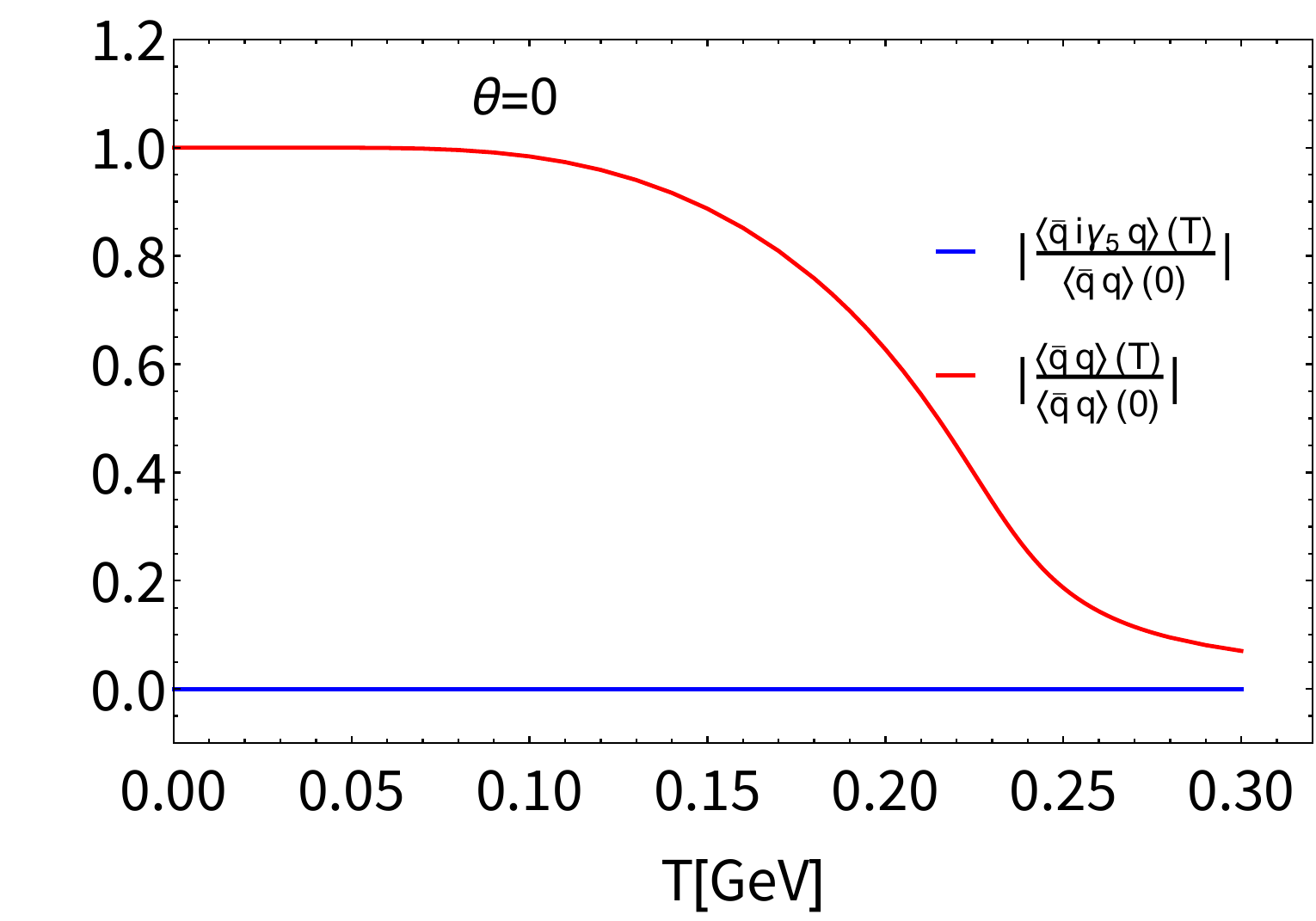}
 \includegraphics[width=0.46\linewidth]{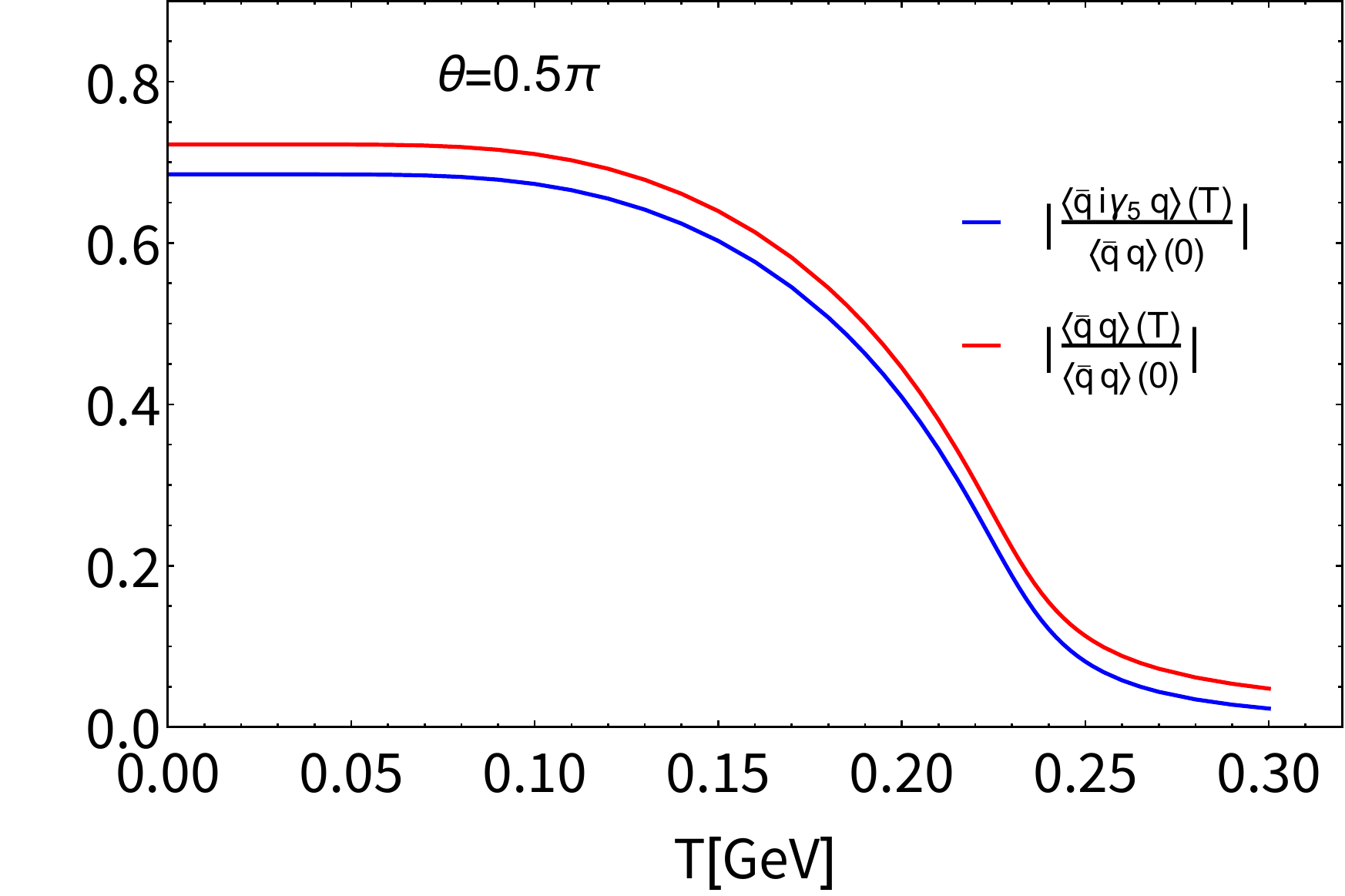}
 \includegraphics[width=0.46\linewidth]{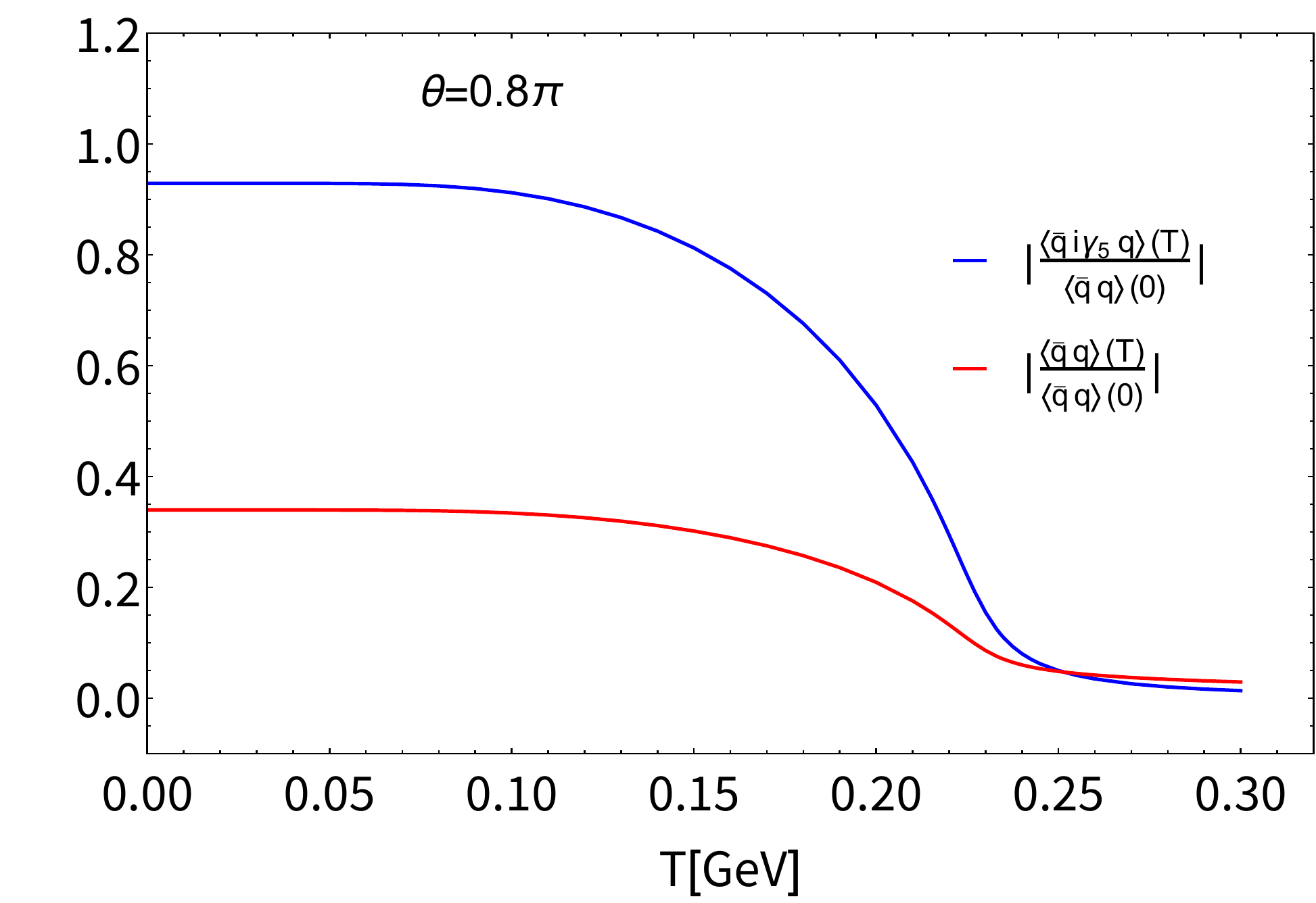}
 \includegraphics[width=0.46\linewidth]{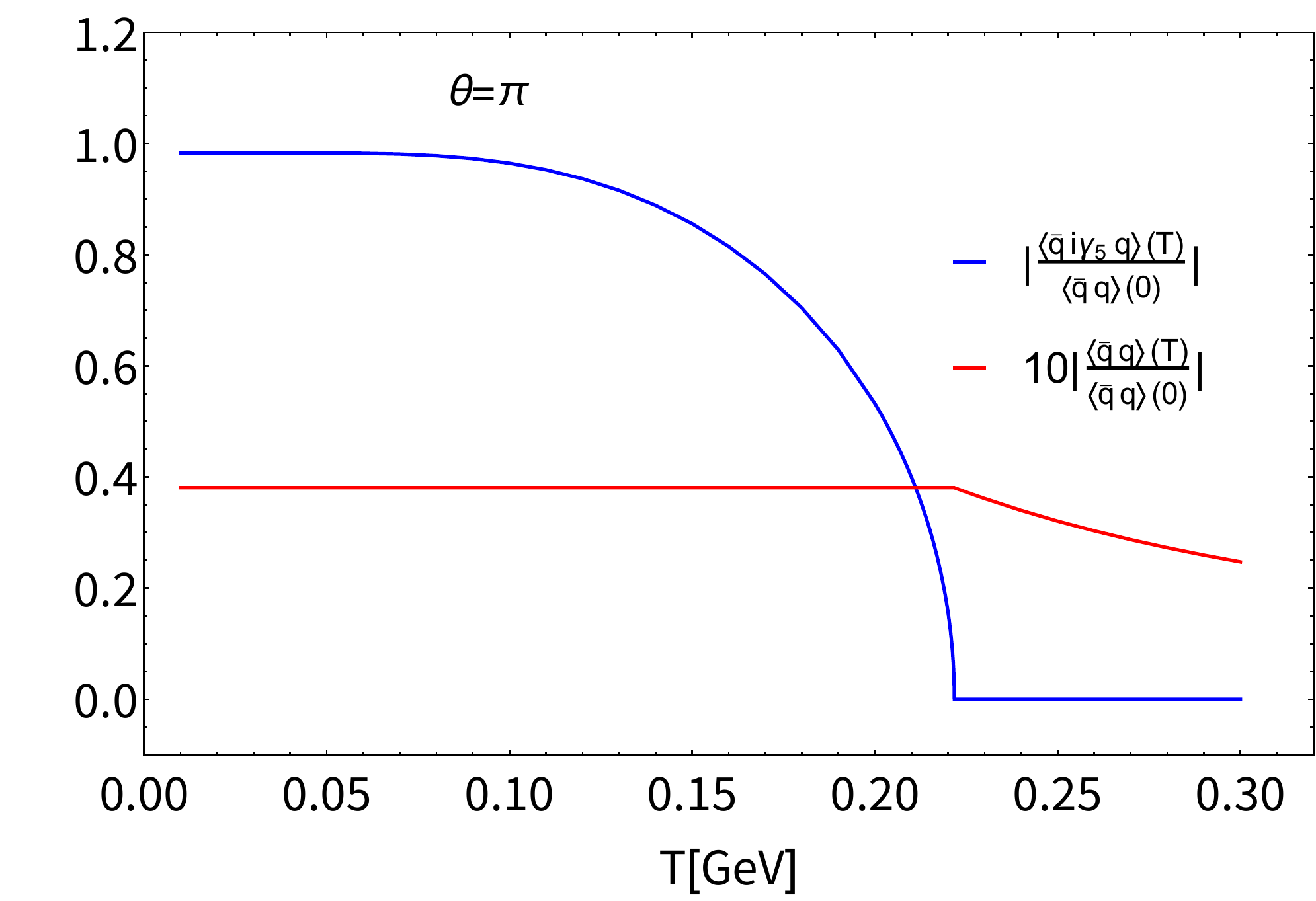}
	\caption{Scalar and pseudoscalar condensates normalized to the scalar condensate at $T = \theta = 0$ versus temperature at nonzero $\theta$. 
 } 
 \label{scalar-pseudo-conden}
\end{figure} 


\textcolor{black}{See Fig.~\ref{scalar-pseudo-conden}, where we observe several characteristic features: 
i) the scalar condensate becomes dramatically smaller with increasing $\theta$, so that $\chi_{\rm top}$ in Eq.(\ref{chitop}) gigantically decreases; 
ii) at $\theta=\pi$, the pseudoscalar condensate undergoes the CP symmetry restoration of the second order type, which generates a significant spike at the criticality in $\chi_{\rm top}$;  
iii) at $\theta=\pi$, 
the scalar condensate does not evolve in $T$ at all until the 
the CP symmetry is restored, in agreement with the analytic discussion around Eq.(\ref{pi-no-chiralSSB}). }

\subsection{PNJL case}

By extending the NJL detailed in the previous subsection into 
the PNJL model in the MFA following the procedure in the literature~\cite{Fukushima:2003fw} (for a recent review, see, e.g., \cite{Fukushima:2017csk}), 
we have the thermodynamic potential: 
\begin{align}
    \Omega_{\rm PNJL}[S', P', L, L^\dag]  
    &= 
    2 g_s (S'^2 + P'^2) + 2g_d (S'^2 - P'^2) 
\notag\\ 
& - 2 N_f \int \frac{d^3 k}{(2 \pi)^3} {\rm tr}_{\rm color} \left[ E + T \left(  
 \ln(1+ L \cdot e^{-E/T})  
 + {\rm h.c.}  \right) \right] 
 + {\cal U}(L, L^\dag) 
 \notag\\ 
 &= 
 2 g_s (S'^2 + P'^2) + 2g_d (S'^2 - P'^2) 
\notag\\ 
& - 2 N_c N_f \int \frac{d^3 k}{(2 \pi)^3} \left[ E + \frac{T}{3} \left( 
 \ln(1+ 3 \Phi e^{-E/T} + 3 \Phi^* e^{-2 E/T} + e^{-3 E/T})   
 +  {\rm h.c.} \right) \right] 
 + {\cal U}(\Phi, \Phi^*) 
 \,, \label{PNJL-Omega}
\end{align}
where  
$\Phi = {\rm tr}[L]/3$,  
in which we have introduced the Polyakov loop field $L = {\rm diag}\{e^{i \phi_1}, e^{i \phi_2}, e^{i \phi_3}  \}$ with the $Z_3$ charges $\phi_{1,2,3}$;  
the $SU(3)$ constraint 
${\rm det}[L]=1$ taken into account; $E$ takes the same form as in Eq.(\ref{M}), 
but would depend on the Polyakov-loop fields $\Phi$ and $\Phi^*$ through 
the stationary condition as in Eq.(\ref{stationary}). 
The Polyakov loop potential ${\cal U}(\Phi, \Phi^*)$ is assumed to take the form of polynomial type or 
logarithmic type:    
\begin{align}
    {\cal U}_{\rm poly}(\Phi, \Phi^*) 
    &= T^4\left[-\frac{b_2(T)}{2} \Phi^* \Phi-\frac{b_3}{6}\left(\Phi^{*3}+\Phi^3\right)+\frac{b_4}{4}(\Phi^* \Phi)^2\right]
    \,, \notag \\ 
    {\cal U}_{\rm log}(\Phi, \Phi^*) 
    &= T^4\left[-\frac{c(T)}{2} \Phi^* \Phi+d(T) \ln \left(1-6 \Phi \Phi^*+4\left(\Phi^3+\Phi^{* 3}\right)-3\left(\Phi \Phi^*\right)^2\right)\right]
    \,,
\end{align}
with 
\begin{align}
    b_2(T)&=a_0+a_1\left(\frac{T_0}{T}\right)+a_2\left(\frac{T_0}{T}\right)^2+a_3\left(\frac{T_0}{T}\right)^3\,, \notag\\ 
    c(T) &= c_0+c_1\left(\frac{T_0}{T}\right)+c_2\left(\frac{T_0}{T}\right)^2
     \,, \notag\\ 
    d(T) &= d_3\left(\frac{T_0}{T}\right)^3
    \,.  
\end{align} 
The Polyakov-loop potential parameters are fixed by fitting the lattice data in the pure Yang-Mills theory as~\cite{Roessner:2006xn,Ratti:2005jh} 
\begin{align}
    \begin{array}{|c|c|c|c|c|c|c|c|c|c|| c|}
\hline a_0 & a_1 & a_2 & a_3 & b_3 & b_4&c_0 & c_1 & c_2 & d_3 & T_0 [{\rm MeV}]\\
\hline 6.75 & -1.95 & 2.625 & -7.44 & 0.75 & 7.5&3.51 & -2.47 & 15.2 & -1.75 & 270 \\
\hline
\end{array} 
\,. 
\end{align}
The thermodynamic potential in Eq.(\ref{PNJL-Omega}) is thus minimized also with respect to $\Phi$ and $\Phi^*$, in addition to the stationary condition along $S'$ and $P'$ directions as in the NJL case, Eq.(\ref{SC-NJL}).

In the RPA the following replacement rule for $I_{P'}$ in Eq.(\ref{Is}) due to the Polyakov-loop contribution is applied: 
\begin{align}
    I_{P'} \to I_{P'}(\Phi)=
    - \frac{ N_c}{\pi^2} \int_0^\Lambda dk k^2 \frac{E^2 - \beta^2}{E^3}  
    \Bigg[
    1- \left( \frac{e^{-E/T} (\Phi + 2 \Phi^* e^{-E/T} + e^{- 2 E/T})}{1 + 3 \Phi e^{-E/T} + 3 \Phi^* e^{-2 E/T} + e^{-3 E/T}} +{\rm h.c.}  \right) 
    \Bigg] \,. 
\end{align}
Similarly for $I_{S'}$ an $I_{S'P'}$ in Eq.(\ref{Is}), we have 
\begin{align}
    I_{S'} \to I_{S'}(\Phi) &= 
    - \frac{ N_c}{\pi^2}  \int_0^\Lambda dk k^2 \frac{E^2 - \alpha^2}{E^3}  
    \Bigg[
    1- \left( \frac{e^{-E/T} (\Phi + 2 \Phi^* e^{-E/T} + e^{- 2 E/T})}{1 + 3 \Phi e^{-E/T} + 3 \Phi^* e^{-2 E/T} + e^{-3 E/T}} +{\rm h.c.}  \right) 
    \Bigg] \,, 
    \notag\\ 
    I_{S'P'}  \to I_{S'P'}(\Phi) &= 
     \frac{ N_c}{\pi^2}   \int_0^\Lambda dk k^2 \frac{2 \alpha \beta}{E^3}  
    \Bigg[
    1- \left( \frac{e^{-E/T} (\Phi + 2 \Phi^* e^{-E/T} + e^{- 2 E/T})}{1 + 3 \Phi e^{-E/T} + 3 \Phi^* e^{-2 E/T} + e^{-3 E/T}} +{\rm h.c.}  \right) 
    \Bigg] \,.
\end{align}
Those modified integral functions are precisely reduced back to the ones in Eq.(\ref{Is}) when 
$\Phi=1$ (without confinement).

\begin{figure}[H] 
\centering
	\includegraphics[width=0.46\linewidth]{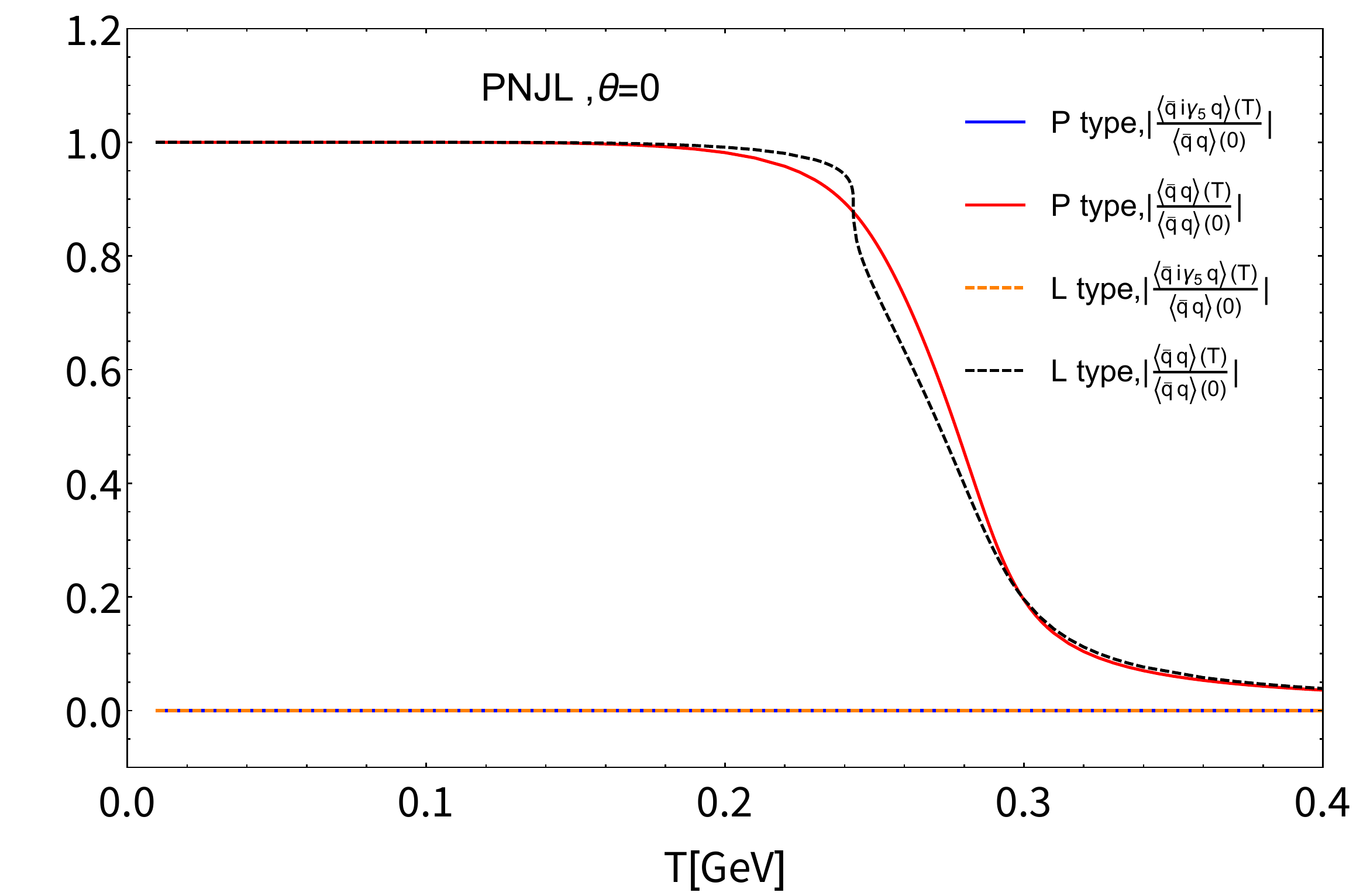}
 \includegraphics[width=0.46\linewidth]{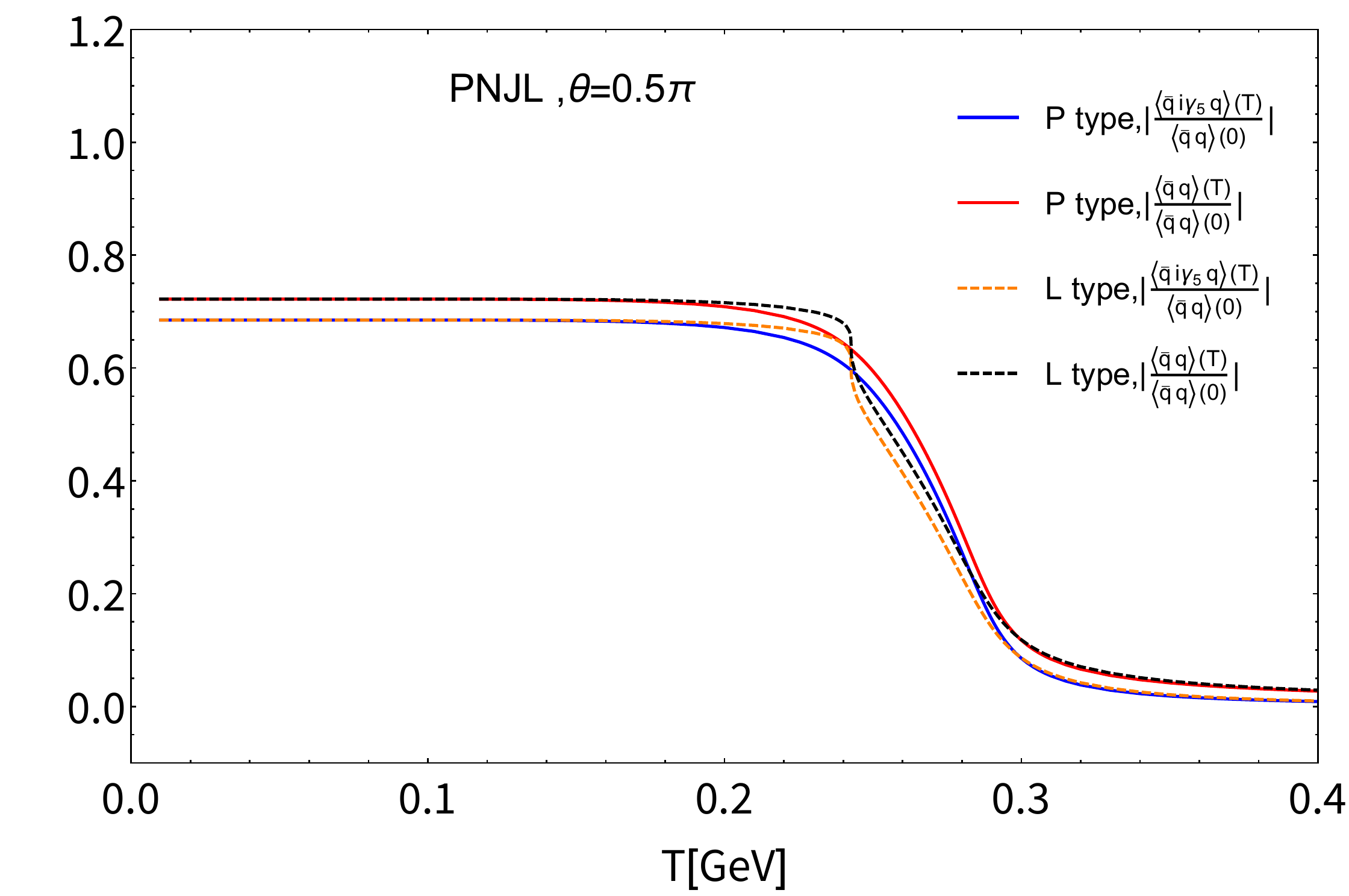}
 \includegraphics[width=0.46\linewidth]{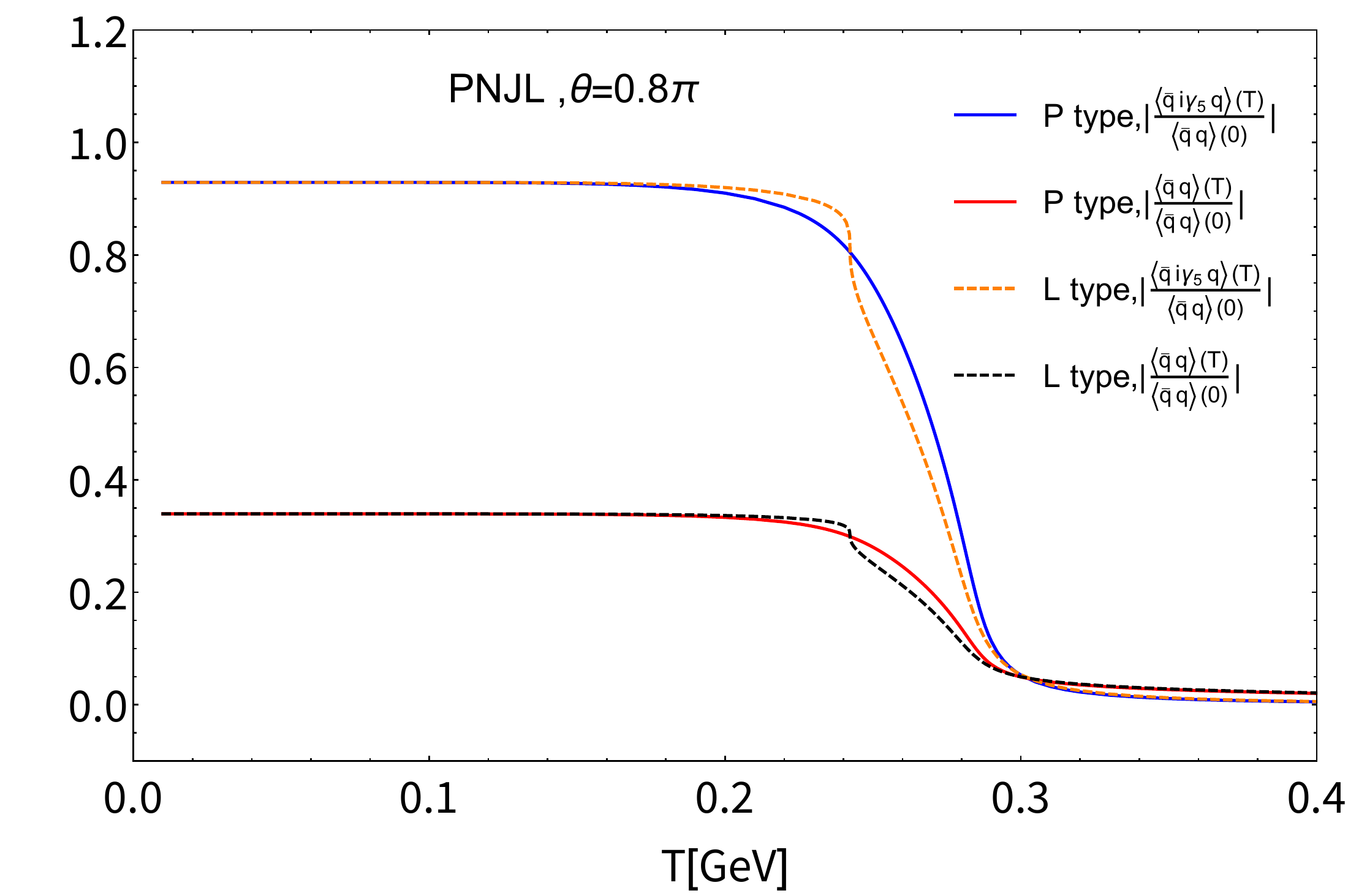}
 \includegraphics[width=0.46\linewidth]{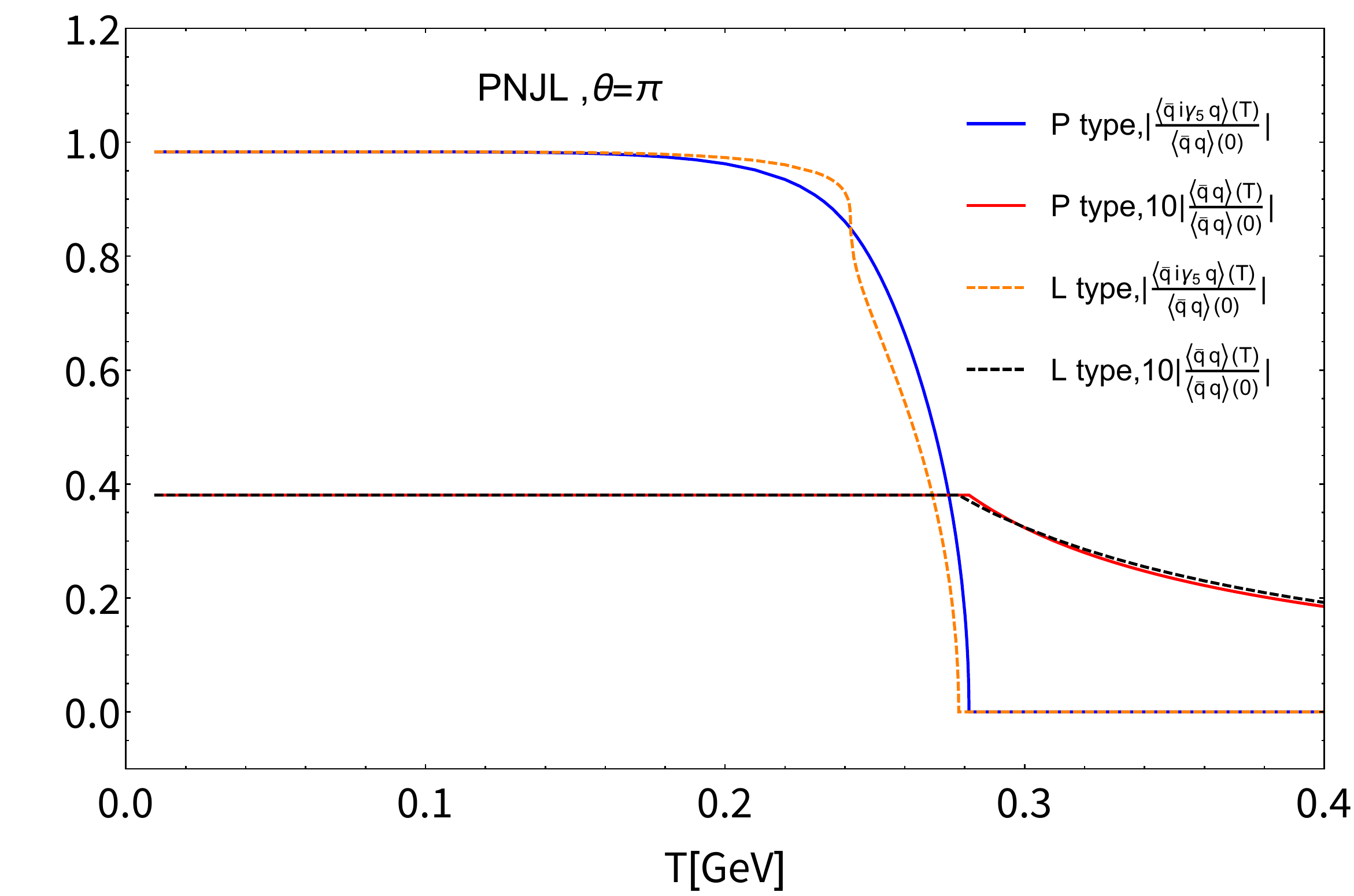}
	\caption{Scalar and pseudoscalar condensates normalized to the scalar condensate 
 at $T=0$ and $\theta=0$ versus temperature with varying $\theta$ from 0. 
 The abbreviations L and P denote the PL potential of logarithmic and polynomial type, respectively. 
 }
 \label{PNJL-scalar-pseudo-conden}
\end{figure}

In Fig.~\ref{PNJL-scalar-pseudo-conden} we plot the scalar and pseudoscalar 
condensates, $\langle \bar{q} q \rangle$ and $\langle \bar{q} i \gamma_5 q \rangle$, 
as a function of temperature $T$ with $\theta/\pi=0, 0.5, 0.8$, and 1.  
The scalar condensate decreases sharply with increasing $\theta$. 
The pseudoscalar condensate undergoes a second-order phase transition at the critical temperature when $\theta=\pi$, and the CP symmetry is restored, which coincides with the NJL case (Fig.~\ref{scalar-pseudo-conden}). 
\textcolor{black}{At $\theta=\pi$, 
the scalar condensate does not evolve in $T$ at all until the CP symmetry is restored at the critical temperature, in the same way as in the NJL case (Fig.~\ref{scalar-pseudo-conden}) due to the same form of the gap equations as in Eq.(\ref{pi-no-chiralSSB}) 
because incorporation of the Polyakov loop field does not break the $U(1)$ axial symmetry. 
This is the characteristic MFA feature, which persists both in the NJL and PNJL cases. }

\begin{figure}[H] 
\centering
	\includegraphics[width=0.48\linewidth]{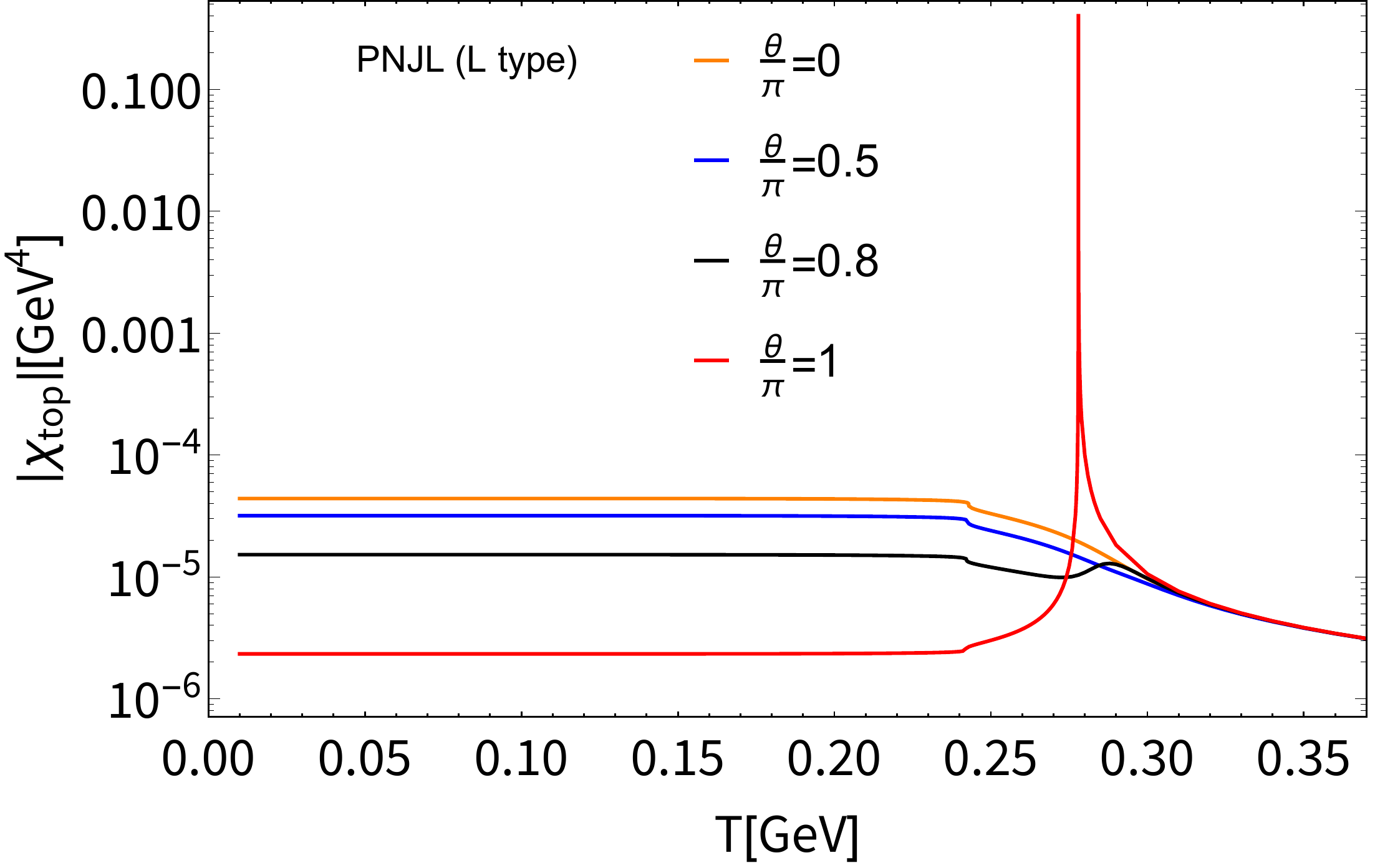}
  \includegraphics[width=0.48\linewidth]{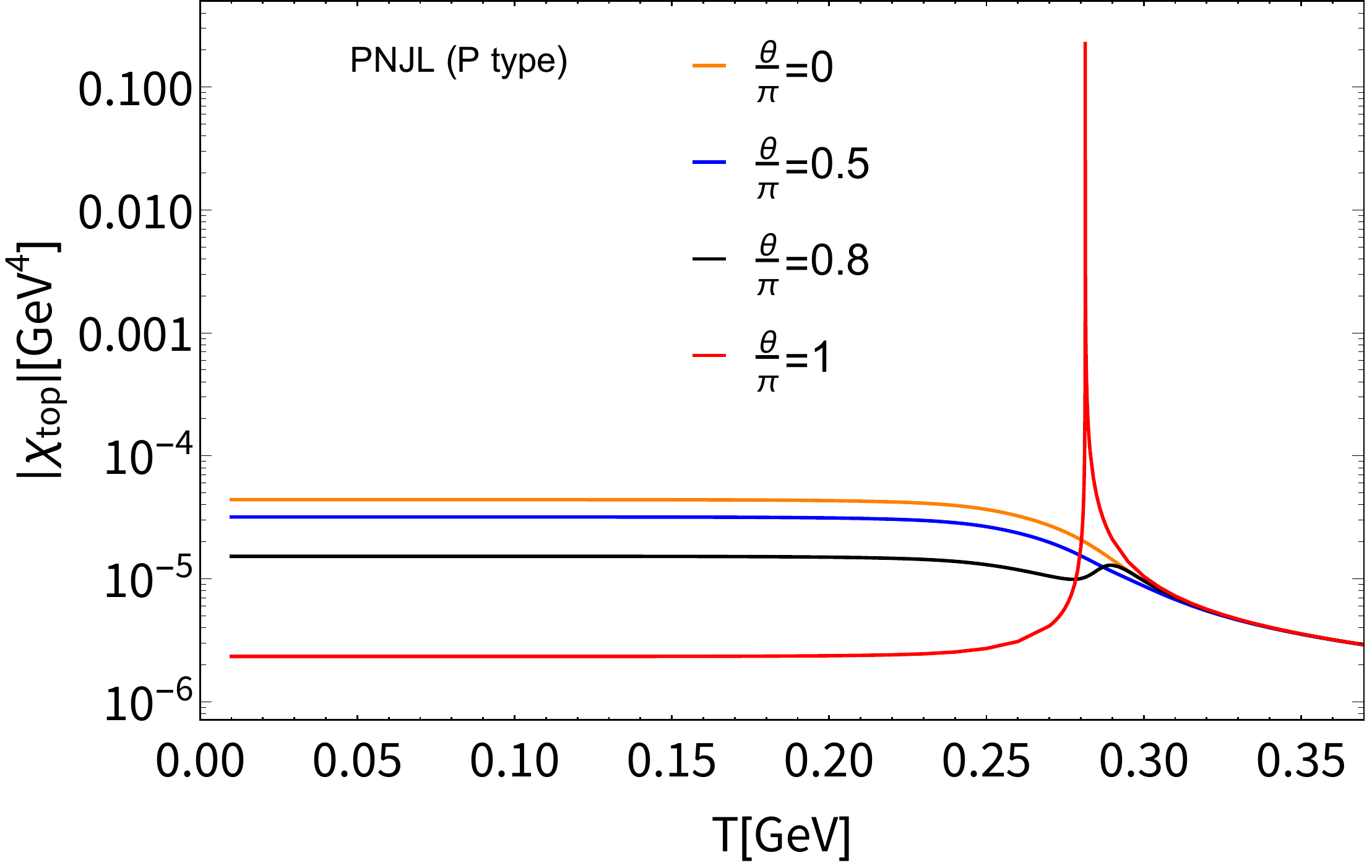}
 \caption{Plot of $\chi_{\rm top}$ (in magnitude) computed from the  two-flavor PNJL model (with two types of PL potentials) with ranging $\theta$ from 0 to $\pi$ 
 as in Fig.~\ref{PNJL-scalar-pseudo-conden}. 
 } 
 \label{PNJL-chi-top}
\end{figure}

\begin{figure}[H] 
\centering
	\includegraphics[width=0.46\linewidth]{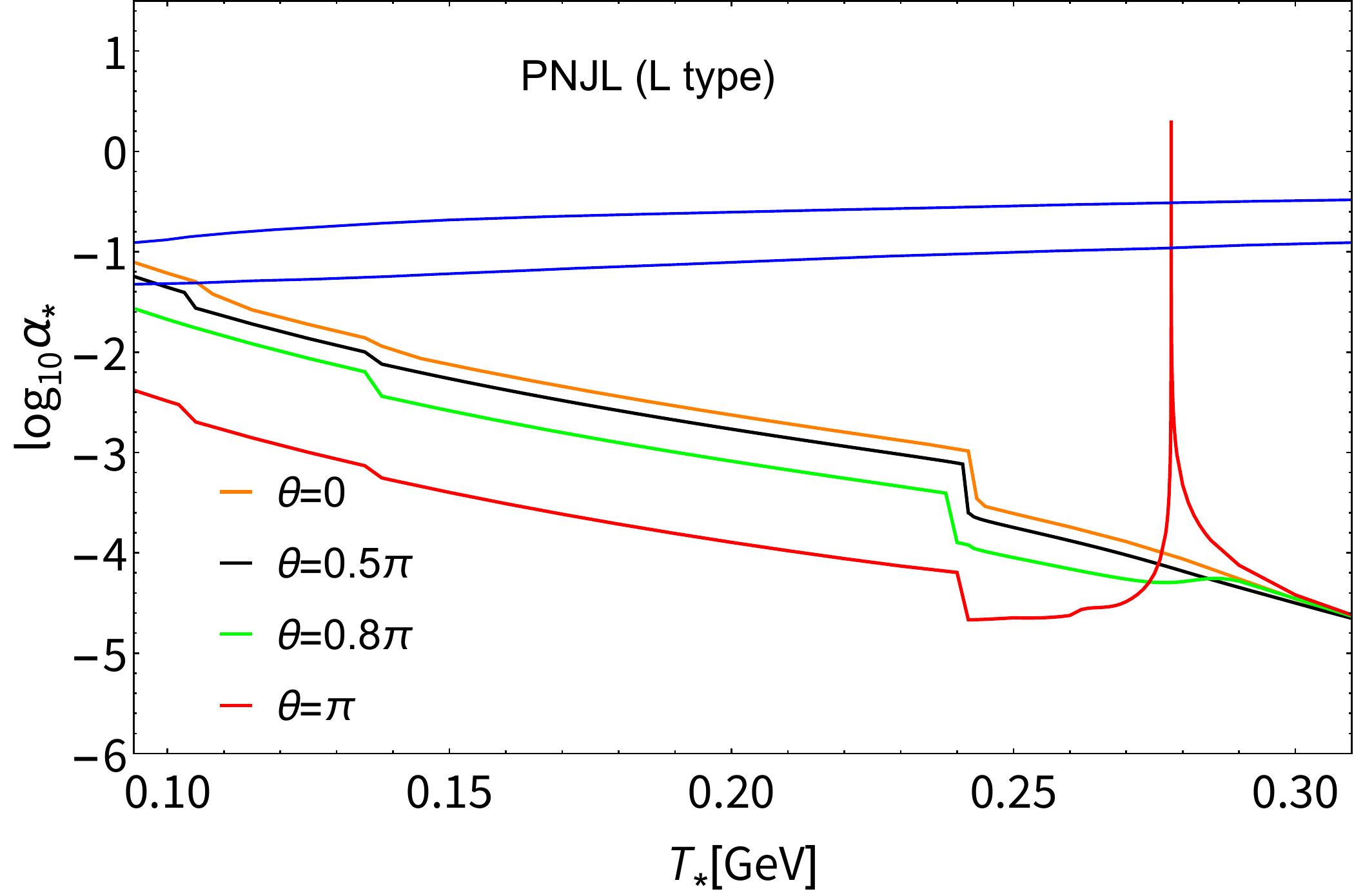}
   \includegraphics[width=0.46\linewidth]{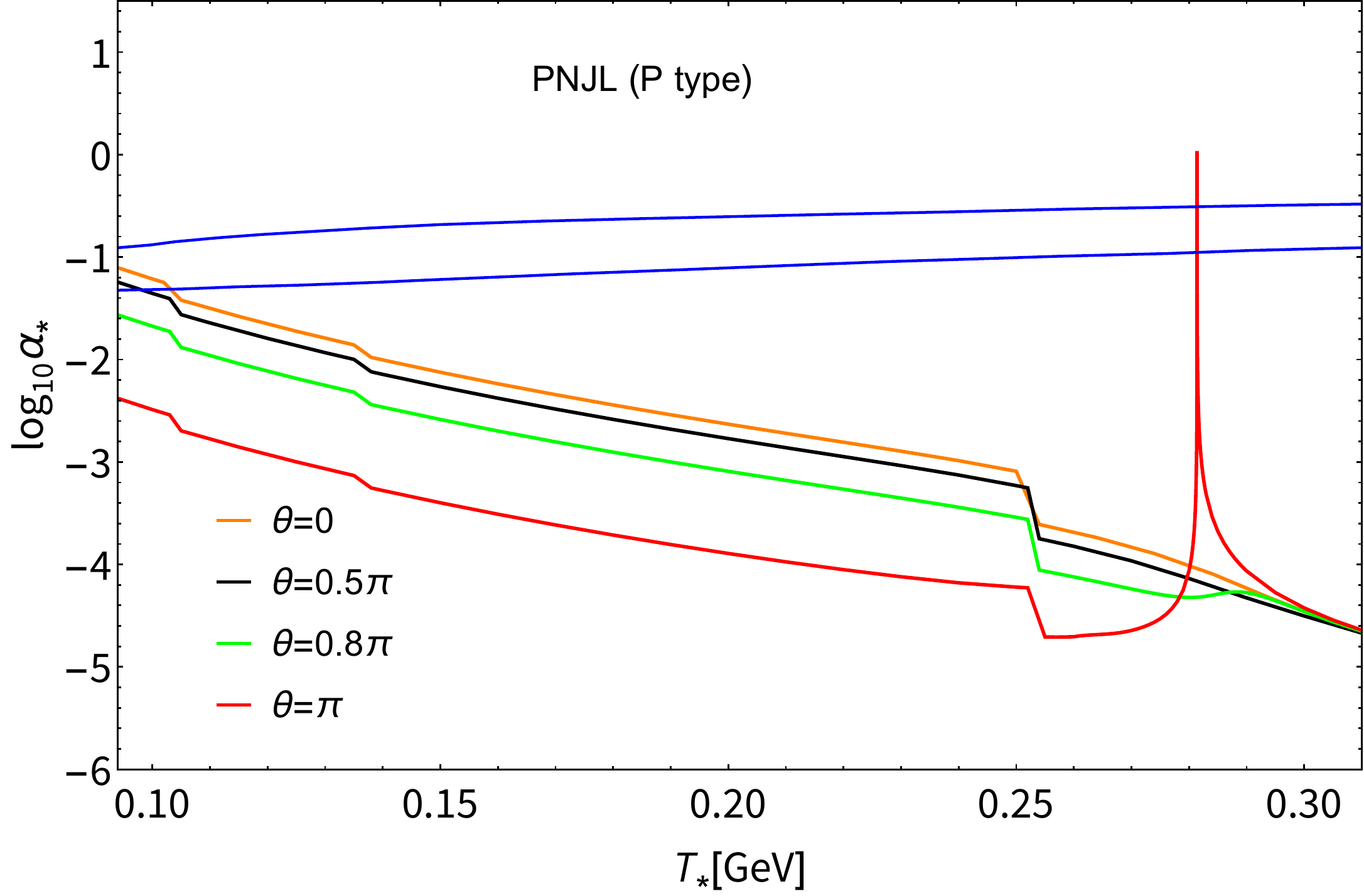}
	\caption{ The signal strength (the normalized latent heat) $\alpha_*$ versus $T^*$ 
 based on the formula in Eq.(\ref{alpha-star}) with 
 the two-flavor PNJL model estimate of $\chi_{\rm top}$ and $\theta$ varied from $0$ to $\pi$ as 
 in Figs.~\ref{PNJL-scalar-pseudo-conden} and \ref{PNJL-chi-top}. 
 } 
 \label{PNJL-alpha}
\end{figure}

Those universal trends have also been seen in $\chi_{\rm top}$, 
in Fig.~\ref{PNJL-chi-top}, which shows no substantial difference from the case of the NJL as in Fig.~\ref{chitop-fig}, 
hence the dramatic suppression of $\alpha_*$ around $T={\cal O}(100)$ MeV still persists for $\theta = {\cal O}(1)$ 
even with the PL contribution taken into account (See Fig.~\ref{PNJL-alpha}).  
At $\theta=\pi$, the peak signal strengths in the PNJL model (whichever type L or P) deviates from the  2$\sigma$ contour, 
as has also been observed in the NJL case (Fig.~\ref{alpha-T-contour-NJL} in the main text).

\begin{thebibliography}{99} 


\bibitem{NANOGrav:2023gor}
G.~Agazie \textit{et al.} [NANOGrav],
Astrophys. J. Lett. \textbf{951}, no.1, L8 (2023)
doi:10.3847/2041-8213/acdac6
[arXiv:2306.16213 [astro-ph.HE]].

\bibitem{NANOGrav:2023hfp}
G.~Agazie \textit{et al.} [NANOGrav],
Astrophys. J. Lett. \textbf{952}, no.2, L37 (2023)
doi:10.3847/2041-8213/ace18b
[arXiv:2306.16220 [astro-ph.HE]].

\bibitem{NANOGrav:2023hvm}
A.~Afzal \textit{et al.} [NANOGrav],
Astrophys. J. Lett. \textbf{951}, no.1, L11 (2023)
doi:10.3847/2041-8213/acdc91
[arXiv:2306.16219 [astro-ph.HE]].

\bibitem{EPTA:2023sfo}
J.~Antoniadis \textit{et al.} [EPTA],
Astron. Astrophys. \textbf{678}, A48 (2023)
doi:10.1051/0004-6361/202346841
[arXiv:2306.16224 [astro-ph.HE]].

\bibitem{EPTA:2023akd}
J.~Antoniadis \textit{et al.} [EPTA and InPTA],
Astron. Astrophys. \textbf{678}, A49 (2023)
doi:10.1051/0004-6361/202346842
[arXiv:2306.16225 [astro-ph.HE]].

\bibitem{EPTA:2023fyk}
J.~Antoniadis \textit{et al.} [EPTA and InPTA:],
Astron. Astrophys. \textbf{678}, A50 (2023)
doi:10.1051/0004-6361/202346844
[arXiv:2306.16214 [astro-ph.HE]].

\bibitem{Reardon:2023gzh}
D.~J.~Reardon, A.~Zic, R.~M.~Shannon, G.~B.~Hobbs, M.~Bailes, V.~Di Marco, A.~Kapur, A.~F.~Rogers, E.~Thrane and J.~Askew, \textit{et al.}
Astrophys. J. Lett. \textbf{951}, no.1, L6 (2023)
doi:10.3847/2041-8213/acdd02
[arXiv:2306.16215 [astro-ph.HE]].

\bibitem{Reardon:2023zen}
D.~J.~Reardon, A.~Zic, R.~M.~Shannon, V.~Di Marco, G.~B.~Hobbs, A.~Kapur, M.~E.~Lower, R.~Mandow, H.~Middleton and M.~T.~Miles, \textit{et al.}
Astrophys. J. Lett. \textbf{951}, no.1, L7 (2023)
doi:10.3847/2041-8213/acdd03
[arXiv:2306.16229 [astro-ph.HE]].

\bibitem{Xu:2023wog}
H.~Xu, S.~Chen, Y.~Guo, J.~Jiang, B.~Wang, J.~Xu, Z.~Xue, R.~N.~Caballero, J.~Yuan and Y.~Xu, \textit{et al.}
Res. Astron. Astrophys. \textbf{23}, no.7, 075024 (2023)
doi:10.1088/1674-4527/acdfa5
[arXiv:2306.16216 [astro-ph.HE]].

\bibitem{Li:2024psa}
H.~J.~Li and Y.~F.~Zhou,
[arXiv:2401.09138 [hep-ph]].

\bibitem{Ellis:2023oxs}
J.~Ellis, M.~Fairbairn, G.~Franciolini, G.~H\"utsi, A.~Iovino, M.~Lewicki, M.~Raidal, J.~Urrutia, V.~Vaskonen and H.~Veerm\"ae,
Phys. Rev. D \textbf{109}, no.2, 023522 (2024)
doi:10.1103/PhysRevD.109.023522
[arXiv:2308.08546 [astro-ph.CO]].

\bibitem{Lozanov:2023rcd}
K.~D.~Lozanov, S.~Pi, M.~Sasaki, V.~Takhistov and A.~Wang,
[arXiv:2310.03594 [astro-ph.CO]].

\bibitem{Gelmini:2023kvo}
G.~B.~Gelmini and J.~Hyman,
Phys. Lett. B \textbf{848}, 138356 (2024)
doi:10.1016/j.physletb.2023.138356
[arXiv:2307.07665 [hep-ph]].

\bibitem{Kitajima:2023cek}
N.~Kitajima, J.~Lee, K.~Murai, F.~Takahashi and W.~Yin,
Phys. Lett. B \textbf{851}, 138586 (2024)
doi:10.1016/j.physletb.2024.138586
[arXiv:2306.17146 [hep-ph]].

\bibitem{Geller:2023shn}
M.~Geller, S.~Ghosh, S.~Lu and Y.~Tsai,
Phys. Rev. D \textbf{109}, no.6, 063537 (2024)
doi:10.1103/PhysRevD.109.063537
[arXiv:2307.03724 [hep-ph]].

\bibitem{Bai:2023cqj}
Y.~Bai, T.~K.~Chen and M.~Korwar,
JHEP \textbf{12}, 194 (2023)
doi:10.1007/JHEP12(2023)194
[arXiv:2306.17160 [hep-ph]].

\bibitem{Blasi:2023sej}
S.~Blasi, A.~Mariotti, A.~Rase and A.~Sevrin,
JHEP \textbf{11}, 169 (2023)
doi:10.1007/JHEP11(2023)169
[arXiv:2306.17830 [hep-ph]].

\bibitem{Borsanyi:2016ksw}
S.~Borsanyi, Z.~Fodor, J.~Guenther, K.~H.~Kampert, S.~D.~Katz, T.~Kawanai, T.~G.~Kovacs, S.~W.~Mages, A.~Pasztor and F.~Pittler, \textit{et al.}
Nature \textbf{539}, no.7627, 69-71 (2016)
doi:10.1038/nature20115
[arXiv:1606.07494 [hep-lat]].

\bibitem{Manton:1983nd}
N.~S.~Manton,
Phys. Rev. D \textbf{28}, 2019 (1983)
doi:10.1103/PhysRevD.28.2019

\bibitem{Klinkhamer:1984di}
F.~R.~Klinkhamer and N.~S.~Manton,
Phys. Rev. D \textbf{30}, 2212 (1984)
doi:10.1103/PhysRevD.30.2212

\bibitem{Kharzeev:2007tn}
D.~Kharzeev and A.~Zhitnitsky,
Nucl. Phys. A \textbf{797}, 67-79 (2007)
doi:10.1016/j.nuclphysa.2007.10.001
[arXiv:0706.1026 [hep-ph]].

\bibitem{Kharzeev:2007jp}
D.~E.~Kharzeev, L.~D.~McLerran and H.~J.~Warringa,
Nucl. Phys. A \textbf{803}, 227-253 (2008)
doi:10.1016/j.nuclphysa.2008.02.298
[arXiv:0711.0950 [hep-ph]].

\bibitem{Fukushima:2008xe}
K.~Fukushima, D.~E.~Kharzeev and H.~J.~Warringa,
Phys. Rev. D \textbf{78}, 074033 (2008)
doi:10.1103/PhysRevD.78.074033
[arXiv:0808.3382 [hep-ph]].

\bibitem{McLerran:1990de}
L.~D.~McLerran, E.~Mottola and M.~E.~Shaposhnikov,
Phys. Rev. D \textbf{43}, 2027-2035 (1991)
doi:10.1103/PhysRevD.43.2027

\bibitem{Moore:1997im}
G.~D.~Moore,
Phys. Lett. B \textbf{412}, 359-370 (1997)
doi:10.1016/S0370-2693(97)01046-0
[arXiv:hep-ph/9705248 [hep-ph]].

\bibitem{Moore:1999fs}
G.~D.~Moore and K.~Rummukainen,
Phys. Rev. D \textbf{61}, 105008 (2000)
doi:10.1103/PhysRevD.61.105008
[arXiv:hep-ph/9906259 [hep-ph]].

\bibitem{Bodeker:1999gx}
D.~Bodeker, G.~D.~Moore and K.~Rummukainen,
Phys. Rev. D \textbf{61}, 056003 (2000)
doi:10.1103/PhysRevD.61.056003
[arXiv:hep-ph/9907545 [hep-ph]].

\bibitem{Andrianov:2012hq}
A.~A.~Andrianov, V.~A.~Andrianov, D.~Espriu and X.~Planells,
Phys. Lett. B \textbf{710}, 230-235 (2012)
doi:10.1016/j.physletb.2012.02.072
[arXiv:1201.3485 [hep-ph]].

\bibitem{Andrianov:2012dj}
A.~A.~Andrianov, D.~Espriu and X.~Planells,
Eur. Phys. J. C \textbf{73}, no.1, 2294 (2013)
doi:10.1140/epjc/s10052-013-2294-0
[arXiv:1210.7712 [hep-ph]].

\bibitem{Kawaguchi:2020qvg}
M.~Kawaguchi, S.~Matsuzaki and A.~Tomiya,
Phys. Rev. D \textbf{103}, no.5, 054034 (2021)
doi:10.1103/PhysRevD.103.054034
[arXiv:2005.07003 [hep-ph]].

\bibitem{Cui:2021bqf}
C.~X.~Cui, J.~Y.~Li, S.~Matsuzaki, M.~Kawaguchi and A.~Tomiya,
Phys. Rev. D \textbf{105}, no.11, 114031 (2022)
doi:10.1103/PhysRevD.105.114031
[arXiv:2106.05674 [hep-ph]].

\bibitem{Cui:2022vsr}
C.~X.~Cui, J.~Y.~Li, S.~Matsuzaki, M.~Kawaguchi and A.~Tomiya,
Particles \textbf{7}, no.1, 237-263 (2024)
doi:10.3390/particles7010014
[arXiv:2205.12479 [hep-ph]].

\bibitem{Leutwyler:1992yt}
H.~Leutwyler and A.~V.~Smilga,
Phys. Rev. D \textbf{46}, 5607-5632 (1992)
doi:10.1103/PhysRevD.46.5607

\bibitem{Kawaguchi:2020kdl}
M.~Kawaguchi, S.~Matsuzaki and A.~Tomiya,
Phys. Lett. B \textbf{813}, 136044 (2021)
doi:10.1016/j.physletb.2020.136044
[arXiv:2003.11375 [hep-ph]].

\bibitem{Dashen:1970et}
R.~F.~Dashen,
Phys. Rev. D \textbf{3}, 1879-1889 (1971)
doi:10.1103/PhysRevD.3.1879

\bibitem{Witten:1980sp}
E.~Witten,
Annals Phys. \textbf{128}, 363 (1980)
doi:10.1016/0003-4916(80)90325-5

\bibitem{Pisarski:1996ne}
R.~D.~Pisarski,
Phys. Rev. Lett. \textbf{76}, 3084-3087 (1996)
doi:10.1103/PhysRevLett.76.3084
[arXiv:hep-ph/9601316 [hep-ph]].

\bibitem{Creutz:2003xu}
M.~Creutz,
Phys. Rev. Lett. \textbf{92}, 201601 (2004)
doi:10.1103/PhysRevLett.92.201601
[arXiv:hep-lat/0312018 [hep-lat]].

\bibitem{Mizher:2008hf}
A.~J.~Mizher and E.~S.~Fraga,
Nucl. Phys. A \textbf{831}, 91-105 (2009)
doi:10.1016/j.nuclphysa.2009.09.004
[arXiv:0810.5162 [hep-ph]].

\bibitem{Boer:2008ct}
D.~Boer and J.~K.~Boomsma,
Phys. Rev. D \textbf{78}, 054027 (2008)
doi:10.1103/PhysRevD.78.054027
[arXiv:0806.1669 [hep-ph]].

\bibitem{Creutz:2009kx}
M.~Creutz,
Annals Phys. \textbf{324}, 1573-1584 (2009)
doi:10.1016/j.aop.2009.01.005
[arXiv:0901.0150 [hep-ph]].

\bibitem{Boomsma:2009eh}
J.~K.~Boomsma and D.~Boer,
Phys. Rev. D \textbf{80}, 034019 (2009)
doi:10.1103/PhysRevD.80.034019
[arXiv:0905.4660 [hep-ph]].

\bibitem{Sakai:2011gs}
Y.~Sakai, H.~Kouno, T.~Sasaki and M.~Yahiro,
Phys. Lett. B \textbf{705}, 349-355 (2011)
doi:10.1016/j.physletb.2011.10.032
[arXiv:1105.0413 [hep-ph]].

\bibitem{Chatterjee:2011yz}
B.~Chatterjee, H.~Mishra and A.~Mishra,
Phys. Rev. D \textbf{85}, 114008 (2012)
doi:10.1103/PhysRevD.85.114008
[arXiv:1111.4061 [hep-ph]].

\bibitem{Sasaki:2011cj}
T.~Sasaki, J.~Takahashi, Y.~Sakai, H.~Kouno and M.~Yahiro,
Phys. Rev. D \textbf{85}, 056009 (2012)
doi:10.1103/PhysRevD.85.056009
[arXiv:1112.6086 [hep-ph]].

\bibitem{Sasaki:2013ewa}
T.~Sasaki, H.~Kouno and M.~Yahiro,
Phys. Rev. D \textbf{87}, no.5, 056003 (2013)
doi:10.1103/PhysRevD.87.056003
[arXiv:1208.0375 [hep-ph]].

\bibitem{Aoki:2014moa}
S.~Aoki and M.~Creutz,
Phys. Rev. Lett. \textbf{112}, no.14, 141603 (2014)
doi:10.1103/PhysRevLett.112.141603
[arXiv:1402.1837 [hep-lat]].

\bibitem{Mameda:2014cxa}
K.~Mameda,
Nucl. Phys. B \textbf{889}, 712-726 (2014)
doi:10.1016/j.nuclphysb.2014.11.002
[arXiv:1408.1189 [hep-ph]].

\bibitem{Verbaarschot:2014upa}
J.~J.~M.~Verbaarschot and T.~Wettig,
Phys. Rev. D \textbf{90}, no.11, 116004 (2014)
doi:10.1103/PhysRevD.90.116004
[arXiv:1407.8393 [hep-th]].

\bibitem{Gaiotto:2017tne}
D.~Gaiotto, Z.~Komargodski and N.~Seiberg,
JHEP \textbf{01}, 110 (2018)
doi:10.1007/JHEP01(2018)110
[arXiv:1708.06806 [hep-th]].

\bibitem{Gaiotto:2017yup}
D.~Gaiotto, A.~Kapustin, Z.~Komargodski and N.~Seiberg,
JHEP \textbf{05}, 091 (2017)
doi:10.1007/JHEP05(2017)091
[arXiv:1703.00501 [hep-th]].

\bibitem{tHooft:1979rat}
G.~'t Hooft,
NATO Sci. Ser. B \textbf{59}, 135-157 (1980)
doi:10.1007/978-1-4684-7571-5\_9

\bibitem{Frishman:1980dq}
Y.~Frishman, A.~Schwimmer, T.~Banks and S.~Yankielowicz,
Nucl. Phys. B \textbf{177}, 157-171 (1981)
doi:10.1016/0550-3213(81)90268-6

\bibitem{Saikawa:2017hiv}
K.~Saikawa,
Universe \textbf{3}, no.2, 40 (2017)
doi:10.3390/universe3020040
[arXiv:1703.02576 [hep-ph]].

\bibitem{Hiramatsu:2010yz}
T.~Hiramatsu, M.~Kawasaki and K.~Saikawa,
JCAP \textbf{05}, 032 (2010)
doi:10.1088/1475-7516/2010/05/032
[arXiv:1002.1555 [astro-ph.CO]].

\bibitem{Hiramatsu:2012sc}
T.~Hiramatsu, M.~Kawasaki, K.~Saikawa and T.~Sekiguchi,
JCAP \textbf{01}, 001 (2013)
doi:10.1088/1475-7516/2013/01/001
[arXiv:1207.3166 [hep-ph]].

\bibitem{Hiramatsu:2013qaa}
T.~Hiramatsu, M.~Kawasaki and K.~Saikawa,
JCAP \textbf{02}, 031 (2014)
doi:10.1088/1475-7516/2014/02/031
[arXiv:1309.5001 [astro-ph.CO]].

\bibitem{ParticleDataGroup:2022pth}
R.~L.~Workman \textit{et al.} [Particle Data Group],
PTEP \textbf{2022}, 083C01 (2022)
doi:10.1093/ptep/ptac097

\bibitem{Aoki:2009sc}
Y.~Aoki, S.~Borsanyi, S.~Durr, Z.~Fodor, S.~D.~Katz, S.~Krieg and K.~K.~Szabo,
JHEP \textbf{06}, 088 (2009)
doi:10.1088/1126-6708/2009/06/088
[arXiv:0903.4155 [hep-lat]].

\bibitem{Borsanyi:2011bn}
S.~Borsanyi \textit{et al.} [Wuppertal-Budapest],
J. Phys. Conf. Ser. \textbf{316}, 012020 (2011)
doi:10.1088/1742-6596/316/1/012020
[arXiv:1109.5032 [hep-lat]].

\bibitem{Ding:2015ona}
H.~T.~Ding, F.~Karsch and S.~Mukherjee,
Int. J. Mod. Phys. E \textbf{24}, no.10, 1530007 (2015)
doi:10.1142/S0218301315300076
[arXiv:1504.05274 [hep-lat]].

\bibitem{Bazavov:2018mes}
A.~Bazavov \textit{et al.} [HotQCD],
Phys. Lett. B \textbf{795}, 15-21 (2019)
doi:10.1016/j.physletb.2019.05.013
[arXiv:1812.08235 [hep-lat]].

\bibitem{Ding:2020rtq}
H.~T.~Ding,
Nucl. Phys. A \textbf{1005}, 121940 (2021)
doi:10.1016/j.nuclphysa.2020.121940
[arXiv:2002.11957 [hep-lat]].

\bibitem{Fukushima:2003fw}
K.~Fukushima,
Phys. Lett. B \textbf{591}, 277-284 (2004)
doi:10.1016/j.physletb.2004.04.027
[arXiv:hep-ph/0310121 [hep-ph]].

\bibitem{Fukushima:2017csk}
K.~Fukushima and V.~Skokov,
Prog. Part. Nucl. Phys. \textbf{96}, 154-199 (2017)
doi:10.1016/j.ppnp.2017.05.002
[arXiv:1705.00718 [hep-ph]].

\bibitem{Ruggieri:2020qtq}
M.~Ruggieri, M.~N.~Chernodub and Z.~Y.~Lu,
Phys. Rev. D \textbf{102}, no.1, 014031 (2020)
doi:10.1103/PhysRevD.102.014031
[arXiv:2004.09393 [hep-ph]].

\bibitem{Chen:2020syd}
S.~Chen, K.~Fukushima, H.~Nishimura and Y.~Tanizaki,
Phys. Rev. D \textbf{102}, no.3, 034020 (2020)
doi:10.1103/PhysRevD.102.034020
[arXiv:2006.01487 [hep-th]].

\bibitem{Kobayashi:1970ji}
M.~Kobayashi and T.~Maskawa,
Prog. Theor. Phys. \textbf{44}, 1422-1424 (1970)
doi:10.1143/PTP.44.1422

\bibitem{Kobayashi:1971qz}
M.~Kobayashi, H.~Kondo and T.~Maskawa,
Prog. Theor. Phys. \textbf{45}, 1955-1959 (1971)
doi:10.1143/PTP.45.1955

\bibitem{tHooft:1976rip}
G.~'t Hooft,
Phys. Rev. Lett. \textbf{37}, 8-11 (1976)
doi:10.1103/PhysRevLett.37.8

\bibitem{tHooft:1976snw}
G.~'t Hooft,
Phys. Rev. D \textbf{14}, 3432-3450 (1976)
[erratum: Phys. Rev. D \textbf{18}, 2199 (1978)]
doi:10.1103/PhysRevD.14.3432

\bibitem{Fejos:2016hbp}
G.~Fejos and A.~Hosaka,
Phys. Rev. D \textbf{94}, no.3, 036005 (2016)
doi:10.1103/PhysRevD.94.036005
[arXiv:1604.05982 [hep-ph]].

\bibitem{Fejos:2023lvw}
G.~Fejos and A.~Patkos,
Phys. Rev. D \textbf{109}, no.3, 036035 (2024)
doi:10.1103/PhysRevD.109.036035
[arXiv:2311.02186 [hep-ph]].

\bibitem{Fejos:2021yod}
G.~Fej\"os and A.~Patkos,
Phys. Rev. D \textbf{105}, no.9, 096007 (2022)
doi:10.1103/PhysRevD.105.096007
[arXiv:2112.14903 [hep-ph]].

\bibitem{DelDebbio:2002xa}
L.~Del Debbio, H.~Panagopoulos and E.~Vicari,
JHEP \textbf{08}, 044 (2002)
doi:10.1088/1126-6708/2002/08/044
[arXiv:hep-th/0204125 [hep-th]].

\bibitem{DElia:2003zne}
M.~D'Elia,
Nucl. Phys. B \textbf{661}, 139-152 (2003)
doi:10.1016/S0550-3213(03)00311-0
[arXiv:hep-lat/0302007 [hep-lat]].

\bibitem{DelDebbio:2006sbu}
L.~Del Debbio, G.~M.~Manca, H.~Panagopoulos, A.~Skouroupathis and E.~Vicari,
PoS \textbf{LAT2006}, 045 (2006)
doi:10.22323/1.032.0045
[arXiv:hep-th/0610100 [hep-th]].

\bibitem{Giusti:2007tu}
L.~Giusti, S.~Petrarca and B.~Taglienti,
Phys. Rev. D \textbf{76}, 094510 (2007)
doi:10.1103/PhysRevD.76.094510
[arXiv:0705.2352 [hep-th]].

\bibitem{Izubuchi:2007rmy}
T.~Izubuchi, S.~Aoki, K.~Hashimoto, Y.~Nakamura, T.~Sekido and G.~Schierholz,
PoS \textbf{LATTICE2007}, 106 (2007)
doi:10.22323/1.042.0106
[arXiv:0802.1470 [hep-lat]].

\bibitem{Vicari:2008jw}
E.~Vicari and H.~Panagopoulos,
Phys. Rept. \textbf{470}, 93-150 (2009)
doi:10.1016/j.physrep.2008.10.001
[arXiv:0803.1593 [hep-th]].

\bibitem{DElia:2013uaf}
M.~D'Elia and F.~Negro,
Phys. Rev. D \textbf{88}, no.3, 034503 (2013)
doi:10.1103/PhysRevD.88.034503
[arXiv:1306.2919 [hep-lat]].

\bibitem{Hirasawa:2024vns}
M.~Hirasawa, K.~Hatakeyama, M.~Honda, A.~Matsumoto, J.~Nishimura and A.~Yosprakob,
PoS \textbf{LATTICE2023}, 193 (2024)
doi:10.22323/1.453.0193
[arXiv:2401.05726 [hep-lat]].

\bibitem{Kitano:2021jho}
R.~Kitano, R.~Matsudo, N.~Yamada and M.~Yamazaki,
Phys. Lett. B \textbf{822}, 136657 (2021)
doi:10.1016/j.physletb.2021.136657
[arXiv:2102.08784 [hep-lat]].

\bibitem{Byrnes:2002gj}
T.~Byrnes, P.~Sriganesh, R.~J.~Bursill and C.~J.~Hamer,
Nucl. Phys. B Proc. Suppl. \textbf{109}, 202-206 (2002)
doi:10.1016/S0920-5632(02)01416-0
[arXiv:hep-lat/0201007 [hep-lat]].

\bibitem{Funcke:2019zna}
L.~Funcke, K.~Jansen and S.~K\"uhn,
Phys. Rev. D \textbf{101}, no.5, 054507 (2020)
doi:10.1103/PhysRevD.101.054507
[arXiv:1908.00551 [hep-lat]].

\bibitem{Kuramashi:2019cgs}
Y.~Kuramashi and Y.~Yoshimura,
JHEP \textbf{04}, 089 (2020)
doi:10.1007/JHEP04(2020)089
[arXiv:1911.06480 [hep-lat]].

\bibitem{Chakraborty:2020uhf}
B.~Chakraborty, M.~Honda, T.~Izubuchi, Y.~Kikuchi and A.~Tomiya,
Phys. Rev. D \textbf{105}, no.9, 094503 (2022)
doi:10.1103/PhysRevD.105.094503
[arXiv:2001.00485 [hep-lat]].

\bibitem{Honda:2021aum}
M.~Honda, E.~Itou, Y.~Kikuchi, L.~Nagano and T.~Okuda,
Phys. Rev. D \textbf{105}, no.1, 014504 (2022)
doi:10.1103/PhysRevD.105.014504
[arXiv:2105.03276 [hep-lat]].

\bibitem{Honda:2022hyu}
M.~Honda,
doi:10.1142/9789811261633\_0003

\bibitem{GomezNicola:2016ssy}
A.~G\'omez Nicola and J.~Ruiz de Elvira,
JHEP \textbf{03}, 186 (2016)
doi:10.1007/JHEP03(2016)186
[arXiv:1602.01476 [hep-ph]].

\bibitem{Roessner:2006xn}
S.~Roessner, C.~Ratti and W.~Weise,
Phys. Rev. D \textbf{75}, 034007 (2007)
doi:10.1103/PhysRevD.75.034007
[arXiv:hep-ph/0609281 [hep-ph]].

\bibitem{Ratti:2005jh}
C.~Ratti, M.~A.~Thaler and W.~Weise,
Phys. Rev. D \textbf{73}, 014019 (2006)
doi:10.1103/PhysRevD.73.014019
[arXiv:hep-ph/0506234 [hep-ph]].

\end{thebibliography}
\end{document}